\documentclass{emulateapj}

\newcommand{\jkas}{J. Korean Astro. Soc.}
\newcommand{\eas}{EAS Pub. Series}
\newcommand{\newast}{New Astronomy}
\newcommand{\ar}{Astron. Rep.}
\newcommand{\appl}{Appl. Phys. Letters}
\newcommand{\jpb}{J. Phys. B: At. Mol. Opt. Phys.}

\newcommand{\subscript}[1]{\textnormal{\scriptsize{#1}}}
\newcommand{\rstar}{\ensuremath{R_\star}}
\newcommand{\rin}{\ensuremath{R_\subscript{in}}}
\newcommand{\rout}{\ensuremath{R_\subscript{out}}}
\newcommand{\tk}{\ensuremath{T_\subscript{K}}}
\newcommand{\trot}{\ensuremath{T_\subscript{rot}}}

\newcommand{\kms}{km~s$^{-1}$}
\newcommand{\klam}{k$\lambda$}
\newcommand{\jybeam}{Jy~beam$^{-1}$}
\newcommand{\irc}{IRC+10216}

\begin{document}

\title{The maser emitting structure and time variability of the SiS lines $J=14-13$ and $15-14$ in \irc\footnote{Based on CARMA and ALMA Cycle0 observations (projects c0710 and ADS/JAO.ALMA\#2011.0.00229.S). Support for CARMA construction was derived from the states of Illinois, California, and Maryland, the James S. McDonnell Foundation, the Gordon and Betty Moore Foundation, the Kenneth T. and Eileen L. Norris Foundation, the University of Chicago, the Associates of the California Institute of Technology, and the National Science Foundation. CARMA development and operations were supported by the National Science Foundation under a cooperative agreement, and by the CARMA partner universities. ALMA is a partnership of ESO (representing its member states), NSF (USA), and NINS (Japan), together with NRC (Canada) and NSC and ASIAA (Taiwan), in cooperation with the Republic of Chile. The Joint ALMA Observatory is operated by ESO, AUI/NRAO and NAOJ.}}

\shorttitle{The $J=14-13$ and $15-14$ SiS masers towards \irc}

\shortauthors{J. P. Fonfr\'{\i}a et al.}

\author{J. P. Fonfr\'ia\altaffilmark{1,2}}
\author{M. Fern\'andez-L\'opez\altaffilmark{3}}
\author{J. R. Pardo\altaffilmark{1,2}}
\author{M. Ag\'undez\altaffilmark{1,2}}
\author{C. S\'anchez Contreras\altaffilmark{4}}
\author{L. Velilla Prieto\altaffilmark{1,2}}
\author{J. Cernicharo\altaffilmark{1,2}}
\author{M. Santander-Garc\'ia\altaffilmark{5}}
\author{G. Quintana-Lacaci\altaffilmark{1,2}}
\author{A. Castro-Carrizo\altaffilmark{6}}
\author{S. Curiel\altaffilmark{7}}

\affiliation{$^1$Molecular Astrophysics Group, Instituto de Ciencia de Materiales de Madrid, CSIC, C/ Sor Juana In\'es de la Cruz, 3, Cantoblanco, 28049, Madrid (Spain)\\
$^2$Molecular Astrophysics Group, Instituto de F\'isica Fundamental, CSIC, C/ Serrano, 123, 28006, Madrid (Spain)\\
$^3$Instituto Argentino de Radioastronom\'ia, CCT-La Plata (CONICET), C.C.5, 1894, Villa Elisa (Argentina)\\
$^4$Department of Astrophysics, Astrobiology Center (CSIC-INTA), Postal address: ESAC campus, P.O. Box 78, E-28691, Villanueva de la Ca\~nada, Madrid (Spain)\\
$^5$Observatorio Astron\'omico Nacional, OAN-IGN, Alfonso XII, 3, E-28014, Madrid (Spain)\\
$^6$Institut de Radioastronomie Millim\'etrique, 300 Rue de la Piscine, 38406 Saint-Martin d'H\`eres (France)\\
$^7$Departamento de Astrof\'isica Te\'orica, Instituto de Astronom\'ia, Universidad Nacional Aut\'onoma de M\'exico, Ciudad Universitaria, 04510, Mexico City (Mexico)}

\begin{abstract}
We present new high angular resolution interferometer observations of the $v=0$ $J=14-13$ and $15-14$ SiS lines towards IRC+10216, carried out with CARMA and ALMA.
The maps, with angular resolutions of $\simeq 0\farcs25$ and 0\farcs55, reveal (1) an extended, roughly uniform, and weak emission with a size of $\simeq 0\farcs5$, (2) a component elongated approximately along the East-West direction peaking at $\simeq 0\farcs13$ and $0\farcs17$ at both sides of the central star, and (3) two blue- and red-shifted compact components peaking around $0\farcs07$ to the NW of the star.
We have modeled the emission with a 3D radiation transfer code finding that the observations cannot be explained only by thermal emission.
Several maser clumps and one arc-shaped maser feature arranged from 5 to 20\rstar{} from the central star, in addition to a thin shell-like maser structure at $\simeq 13\rstar$ are required to explain the observations.
This maser emitting set of structures accounts for 75\% of the total emission while the other 25\% is produced by thermally excited molecules.
About 60\% of the maser emission comes from the extended emission and the rest from the set of clumps and the arc.
The analysis of a time monitoring of these and other SiS and $^{29}$SiS lines carried out with the IRAM 30~m telescope from 2015 to present suggests that the intensity of some spectral components of the maser emission strongly depends on the stellar pulsation while other components show a mild variability.
This monitoring evidences a significant phase lag of $\simeq 0.2$ between the maser and NIR light-curves.
\end{abstract}


\section{Introduction}
\label{sec:introduction}

Maser emission is commonly produced in the circumstellar envelopes of both O- and C-rich evolved stars \citep*[e.g.,][]{cho_2009}.
O-rich stars mostly display masers of SiO, H$_2$O, and OH \citep*[e.g.,][]{jewell_1987,cernicharo_1993,melnick_1993,yates_1995,humphreys_1997,gray_1999,herpin_1998,pardo_1998,menten_2006,gray_2009,kwon_2012} while HCN masers are usually detected in C-rich stars \citep{guilloteau_1987,lucas_1988,bieging_2000,schilke_2000,bieging_2001,schilke_2003,menten_2018}.
Masers produced in the outer envelope (H$_2$O and OH; with $n_\subscript{gas}\lesssim 5\times 10^4$~cm$^{-3}$) have been proposed to be collisionally pumped at low kinetic temperatures and densities \citep{neufeld_1991,cernicharo_1994,cernicharo_1999,cernicharo_2006a,cernicharo_2006b} although the IR pumping seems to influence significantly the population inversion of the rotational levels \citep{gonzalez-alfonso_1999,he_2004,he_2005}.
Contrarily, the maser emission coming from the inner envelope (SiO and HCN; with $n_\subscript{gas}\gtrsim 10^7$~cm$^{-3}$) seems to be consequence of three pumping mechanisms:
(1) an increase of the trapping lifetime of optically thick ro-vibrational transitions with low $J$ \citep{kwan_1974,bujarrabal_1981,goldsmith_1988,lockett_1992}, 
(2) overlaps between ro-vibrational lines of different molecular species or isotopologues \citep*[e.g.,][]{cernicharo_1991,pardo_1998}, and
(3) changes in the population of rotational levels of different vibrational states due to, e.g., Coriolis resonance favored by the continuum emission in the NIR \citep{lide_1967,schilke_2000,schilke_2003}.
Maser emission uses to come from spots that sample different regions of the envelope  \citep*[e.g.,][]{humphreys_2002,bains_2003,soria-ruiz_2004,soria-ruiz_2007}.
Thus, a detailed analysis of the maser emission, its excitation mechanism, and the structure of the maser emitting regions can provide us with valuable information about the gas dynamics and excitation conditions throughout the circumstellar envelopes \citep*[e.g.,][]{cernicharo_1994,cernicharo_1997}.

The C-rich AGB star \irc, located at $\simeq 120$~pc from Earth \citep{groenewegen_2012}, is surrounded by an optically thick circumstellar envelope composed of dust and molecular gas.
Among the most abundant molecules close to the star are HCN, C$_2$H$_2$, and SiS with abundances with respect to H$_2$ $\gtrsim 5\times 10^{-6}$ \citep{cernicharo_1999,cernicharo_2011,fonfria_2008,fonfria_2015}.

Several decades ago, \citet{henkel_1983} found that the SiS($1-0$) line in the vibrational ground state displays maser emission towards this star, something unobserved in other astronomical objects so far.
This maser was deduced to come from the inner shells of the envelope and it was thought to be pumped due to the reduced escape probability prevailing in this region, as it happens with the SiO $v>0$ masers.
The maser nature of the SiS($1-0$) line of the vibrational ground state has been recently demonstrated by \citet{gong_2017} who propose that the IR continuum is the main responsible of the population inversion.

\citet{fonfria_2006} reported the discovery of new maser emission toward this star in the SiS lines $J=11-10$, $14-13$, and $15-14$ also of the vibrational ground state.
These masers were explained as the consequence of different overlaps in the MIR range between SiS ro-vibrational lines and strong features of C$_2$H$_2$ and HCN.
These overlaps would occur in the region of the circumstellar envelope ranging from the stellar photosphere to $\simeq 20-30\rstar$ (dust formation zone), where $1\rstar\simeq 0\farcs014-0\farcs023$ \citep*[e.g.,][]{ridgway_1988,keady_1988,monnier_2000a,menshchikov_2001}.
\citet{fonfria_2014} interferometer observations of the SiS($14-13$) line with a synthetic Half Power Beam Width (HPBW)~$\simeq 0\farcs25$ and a channel width $\delta v\simeq 12.3$~\kms{} confirming that this line shows maser emission coming mainly either from several isolated clumps or a single inhomogeneous region extending from $\simeq 5$ to 20\rstar.
The monitoring of this line over a period of 2~yr with a single-dish telescope carried out recently by \citet{he_2017} also supports its maser nature after analyzing the variability of their maser peaks.

In this paper, we present and analyze new high angular, high spectral resolution interferometer observations of the SiS($14-13$) maser (HPBW~$\simeq 0\farcs25$, $\delta v\simeq 0.92$~\kms) in \irc{} carried out with the Combined Array for Research in Millimeter-wave Astronomy (CARMA).
Lower spatial resolution observations of the $J=15-14$ line acquired with the Atacama Large Millimeter Array (ALMA; HPBW~$\simeq 0\farcs55$, $\delta v\simeq 0.54$~\kms) are also presented and analyzed.
Part of this analysis is complemented with a new detailed monitoring of the molecular emission during a whole pulsation period carried out with the IRAM 30~m telescope.
We aim to further constrain the structure and excitation conditions of the maser emitting regions as well as its time variability.
We present the observations in Section~\ref{sec:observations}.
The model used to analyze them and the results of this analysis are presented in Sections~\ref{sec:results} and \ref{sec:modeling}.
These results are discussed in Section~\ref{sec:discussion}.

\section{Observations}
\label{sec:observations}

\subsection{CARMA observations}

The CARMA observations were carried out on 2011 January 5 in its 15-antenna mode under the frame of project c0710 \citep{fonfria_2014}.
The antennas were arranged in the B array configuration, with baselines in the range $70-830$~\klam{} and system temperatures of about 200~K.
The weather during the observations was good with an atmospheric opacity $\simeq 0.1$ at 230~GHz.

The calibration and data reduction were performed in the standard way using the Miriad package. 
Bandpass solutions were obtained observing 3C84, while 0854+201 was periodically monitored every 15~min to calibrate phases. 
The observing track went on during more than 5~hrs and we integrated on-source for more than 3~hrs so that the atmospheric variations are expected to be averaged in the final maps.
  The phase calibrator was at an angular distance of $\simeq 15^\circ$ from \irc, which could in principle introduce an error in the phase calibration difficult to quantify although it should be small as the weather was good.
Calibrator 0854+201 was assumed to have a flux density of 3.8~Jy at 257~GHz, as determined by the CARMA and Submillimeter Array (SMA) quasar monitoring programs near the time of our observations.
The absolute flux calibration uncertainty is about 15\%.

The SiS($14-13$) line was observed using a spectral window with a bandwidth of 125~MHz and a spectral resolution of 0.78~MHz ($\simeq 0.92$~\kms{} at the observed frequency).
This observation was taken at the same time than the lower spectral resolution data of this line presented by \citet{fonfria_2014}.
The obtained maps have a synthesized beam of about $0\farcs25$ resolving partially the structure around the central star. 
The primary beam of the telescopes was 30\arcsec.
The dirty PSF show a few significant sidelobes with a peak of $\simeq 30$\% located at $\simeq 1\arcsec$ from the centre of the main beam.
These sidelobes are not expected to significantly affect the shape of the lowest level contours of the brightness distribution considered in this work.

The noise RMS determined from the statistical analysis of a velocity channel with no emission is $\simeq 45$~m\jybeam.
Following the procedure used by \citet{fonfria_2014}, we derive a statistical position uncertainty of $\simeq 2$~mas.
We estimate the systematic uncertainty to be $\simeq 30$~mas after analyzing the RMS of the visibility phases of calibrator 0854+201 and the position of the peak continuum emission of \irc{} in each observing cycle.
Hence, the position uncertainty is clearly dominated by the systematic error.
With this error, the derived stellar position is compatible with the very accurate results by \citet{menten_2012}.

\subsection{ALMA Cycle 0 observations}

The ALMA observations were taken on 2012 April 8 to 23 during Cycle 0 \citep*[project 2011.0.00229.S;][]{cernicharo_2013}.
The array consisted of 16 antennas with baselines from $\simeq 30$ up to 370~\klam.
The weather was good resulting in system temperatures ranging from 100 to 130~K.

The calibration and data reduction were performed in the standard way using the CASA software.
Bandpass and absolute flux were calibrated with observations of 3C279 and 3C273.
J0854+201 and J0909+013 were observed every 10 and 20 minutes to calibrate phase and amplitude.
The absolute flux calibration uncertainty is about 8\%.
See \citet{cernicharo_2013} for a deeper insight into the calibration process.

The observations covered the spectral range $\simeq 270-274$~GHz with a channel width of $\simeq 0.49$~MHz ($\simeq 0.54$~\kms{} at the frequency of the SiS line $J=15-14$) and an effective spectral resolution of $\simeq 1.1$~\kms.
The data were weighted assuming a robust parameter of 0.5, which produced maps with an angular resolution of $\simeq 0\farcs55$.

The noise RMS is $\simeq 5$~m\jybeam.
Following the same method to estimate the position uncertainty than for the CARMA observations, we find that the statistical contribution is negligible.
Again, the systematic uncertainty dominates and the position error can be considered as $\simeq 80$~mas.

\subsection{IRAM 30~m telescope observations}

In this work, we have used the IRAM 30~m telescope observations of the $J=14-13$ and $15-14$ SiS lines carried out on 2004 June 19 and published by \citet{fonfria_2006}.
We also analyzed a new set of SiS (i.e., $^{28}$SiS) and $^{29}$SiS lines recently observed as part of a more ambitious project targeting the time variability of the \irc{} molecular emission.
This monitoring, motivated by the discovery of the time variability of the thermal line emission by \citet{cernicharo_2014}, was carried out from 2015 Jun 7 to present with an average sampling time of 16~days.
We use the wobbler system to get very flat baselines, the EMIR receivers in 7 different tunings, and the BBC, WILMA and FTS backends.
All this allows to completely cover 84~GHz within the $80-273$~GHz range in each observing run.

Pointing and focus are checked every $90-100$~min using the nearby quasar OJ287 (0854+201).
The maximum frequency resolution achieved is 195~kHz in all bands which translates into velocity resolutions from $0.21$ to $0.73$~\kms.
Typical integration times range from $\simeq 20$ to 50~min, which ensures $T_A^*$ noise levels below 20~mK for each individual observation (just a few mK at the lowest frequencies).
The whole data set will be presented and analyzed in \citet{pardo_2018}.

The data imaging of the CARMA and ALMA data sets as well as the single-dish data reduction were achieved using the GILDAS software\footnote{\url{http://www.iram.fr/IRAMFR/GILDAS}}.

\section{Observational results}
\label{sec:results}

\subsection{Spatial distribution of the maser emission}

\begin{figure}
\centering
\includegraphics[width=0.475\textwidth]{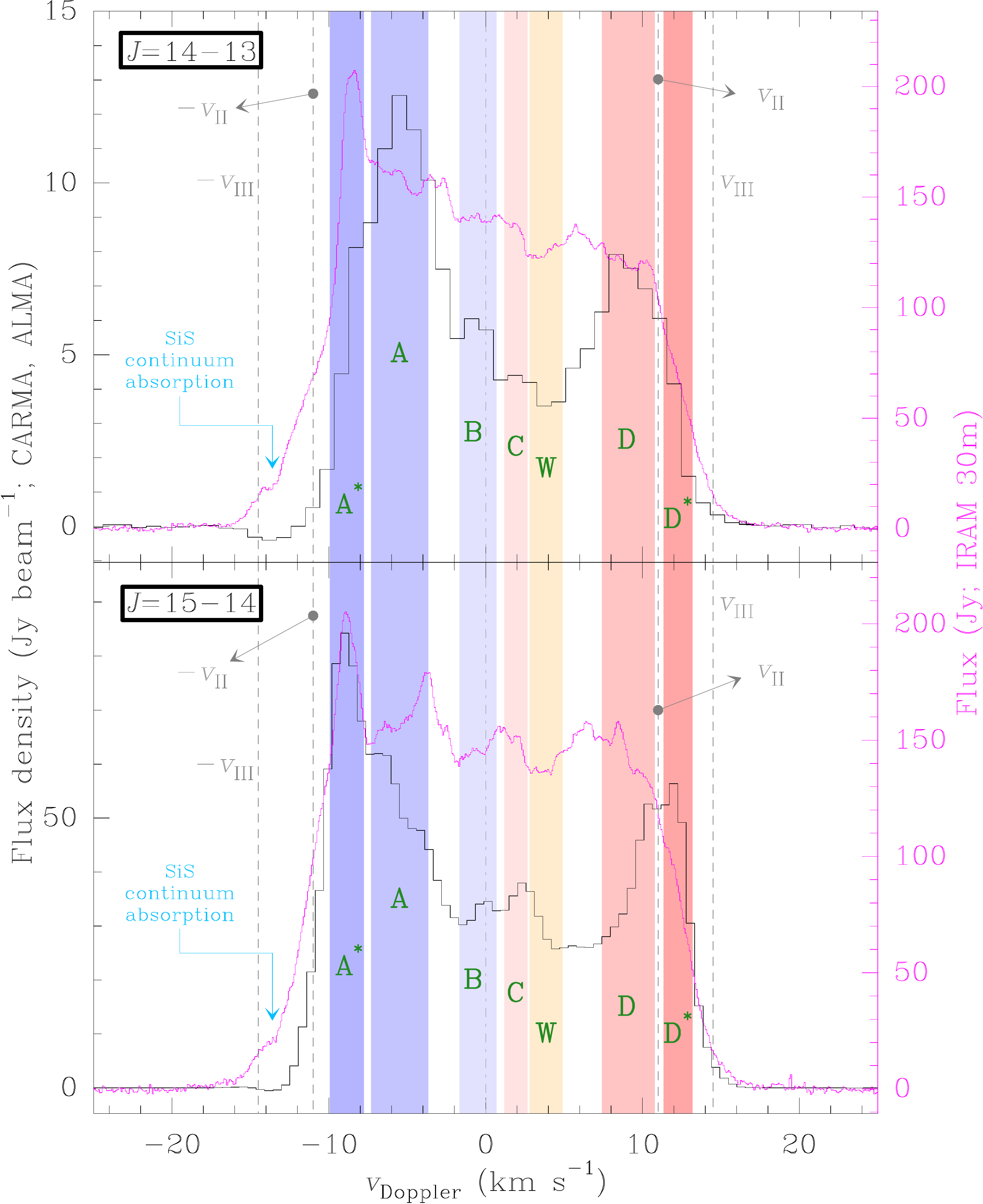}
\caption{Observed spectra at the pixel where the continuum peaks (central pixel) of the SiS lines $J=14-13$ and $15-14$ carried out with CARMA and ALMA (black histograms in the upper and lower panels, respectively).
  The Doppler velocity is referred to the systemic velocity of \irc{} \citep*[$\simeq -26.5$~\kms; e.g.,][]{cernicharo_2000}.
Complementary single-dish data acquired with the IRAM 30~m telescope during 2004 are also shown \citep*[magenta histograms;][]{fonfria_2006}.
$v_\subscript{II}$ and $v_\subscript{III}$ are the gas expansion velocity from 5 to 20\rstar{} and beyond (Section~\ref{sec:modeling}).
The so-called emission components A, B, C, D, and W (green; see Sections~\ref{sec:observations} and \ref{sec:results} to find their spatial counterparts) are produced by a number of clumps of the masing structure.
Components A$^*$ and D$^*$ are not present in the SiS($14-13$) line observed with CARMA (see Sections~\ref{sec:results} and \ref{sec:discussion}).}
\label{fig:f1}
\end{figure}

\begin{figure*}
  \centering
\includegraphics[width=0.9\textwidth]{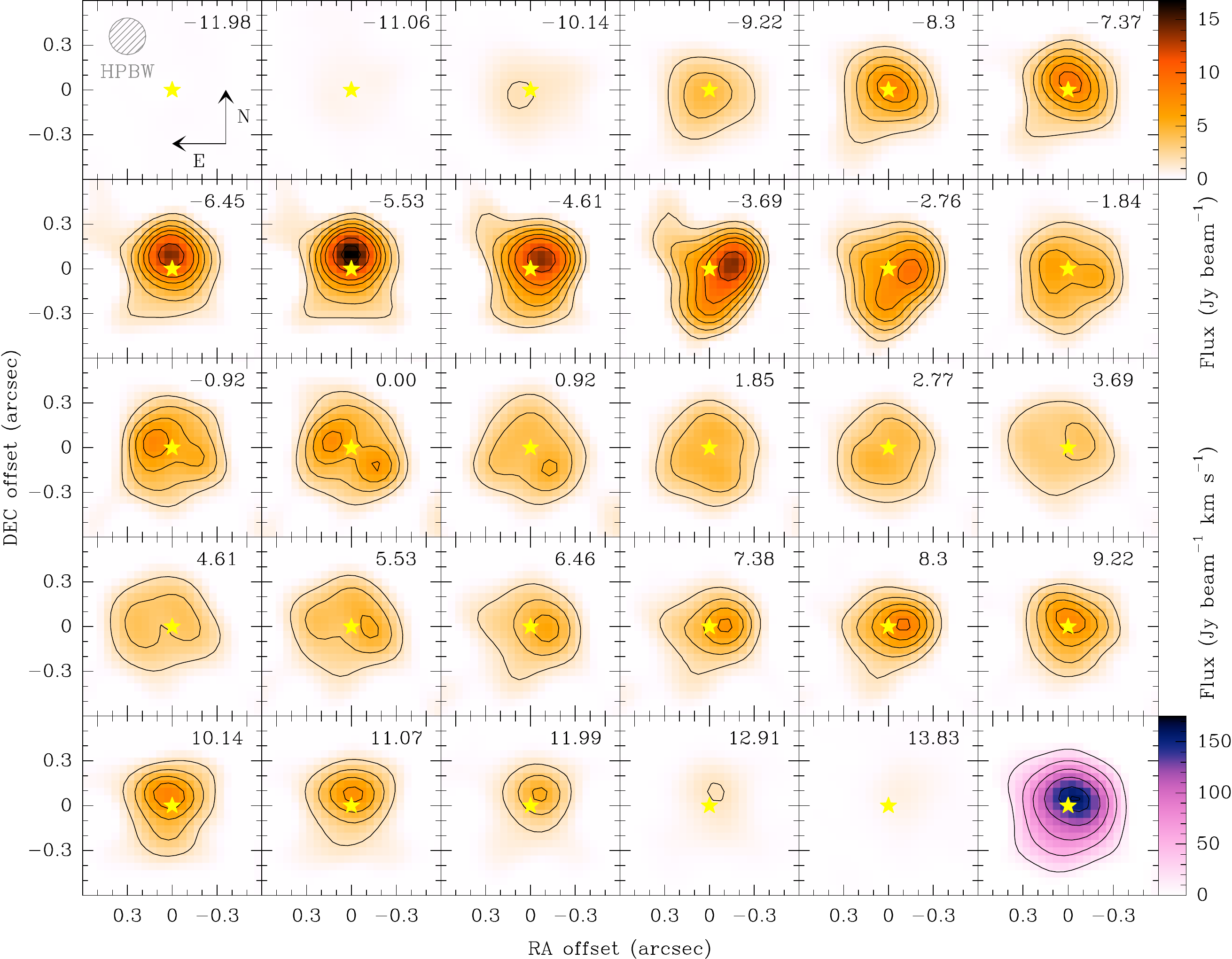}
\caption{Velocity-channel maps of the observed brightness distribution (orange scale).
The distribution of the integrated emission (moment 0 map) is plotted in the bottom right panel (magenta scale).
The contours are at levels 10, 20, 30, 40, 50, 70, 90, and 99\% of the peak emission of the whole cube (velocity-channel maps) and at 10, 30, 50, 70, 90, and 99\% of the total emission (moment 0 map).
The Doppler velocity expressed in \kms{} of each map is close to the top right corner and it is referred to the systemic velocity ($\simeq -26.5$~\kms).
}
\label{fig:f2}
\end{figure*}

\begin{figure*}
  \centering
\includegraphics[width=0.9\textwidth]{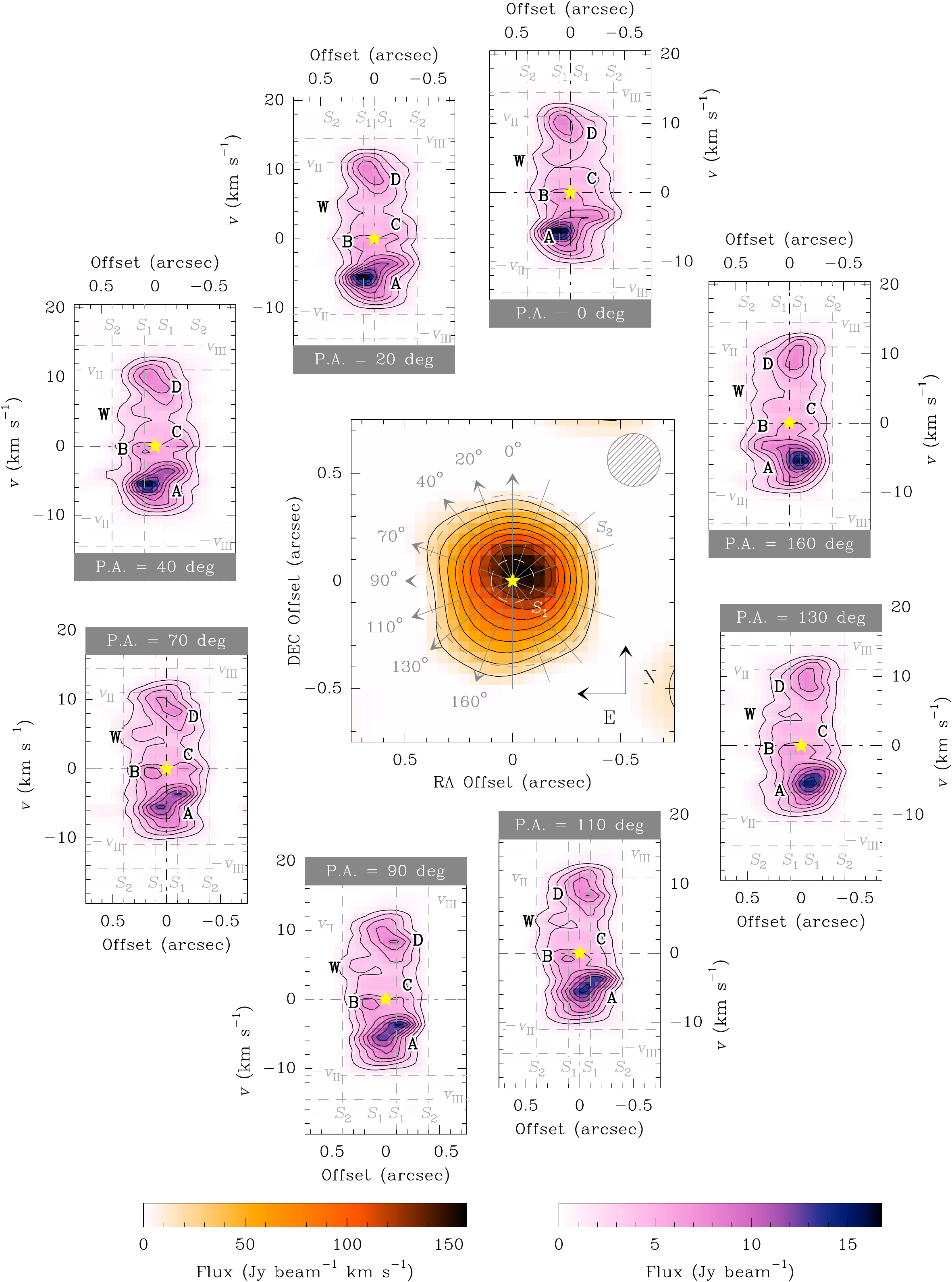}
\caption{Moment 0 map of the \textit{observed} brightness distribution of the SiS($14-13$) line (central panel) and position-velocity maps for the slices with P.A.~=~0, 20, 40, 70, 90, 110, 130, and 160$\degr$.
The contours in black are at levels 10 to 90\% by 10\% of the maximum flux of the moment 0 observed emission ($\simeq 160.3$~\jybeam~\kms) and of the cube ($\simeq 16.9$~\jybeam) for the position-velocity diagrams.
The synthetic PSF is plotted in the panel containing the moment 0 map.
In these diagrams, the main features of the observed emission profile are labelled as A, B, C, D, and W.
The dashed circles and lines $S_1$ and $S_2$ represent the shells at 5 and 20\rstar, and $v_\subscript{II}=11$~\kms{} and $v_\subscript{III}=14.5$~\kms{}
to the gas expansion velocity of Regions II and III (Section~\ref{sec:modeling}).
The velocities are given with respect to the systemic velocity ($\simeq -26.5$~\kms)}.
\label{fig:f3}
\end{figure*}

The comparison between the spectra taken at the central pixel of lines $J=14-13$ and $15-14$ shows many similarities but also prominent differences (Figure~\ref{fig:f1}).
The SiS($14-13$) line is mostly composed of four emission components (A, B, C, and D), a deficit of emission (W; see also Figures~\ref{fig:f2} and \ref{fig:f3}), and a weak blue-shifted absorption mostly noticeable below the continuum in the interferometer observations.
The fingerprints of all these components can be found also in line $J=15-14$ although two new components, named A$^*$ and D$^*$ from now on, are noticeably blue- and red-shifted with respect to components A and D.
It is remarkable that the velocity of the emission peak of line $J=14-13$ does not match up with that of the emission peak shown by the same line observed with the IRAM 30~m telescope contrarily to line $J=15-14$ (ALMA), that shows the same peak velocity than its single-dish counterpart.
It is not clear whether the A and D components comprise the A$^*$ and D$^*$ components or they are just separate components (the ALMA data have no enough angular resolution to spatially disentangle the emission), strongly affected by the time evolution of the envelope.
In addition to the easily visible component A$^*$, hints of components A, B, C, D, D$^*$, W, and the blue-shifted absorption can also be found in the single-dish observations acquired with the IRAM 30~m telescope reported by \citet{fonfria_2006} and the new data set taken at different stellar pulsation phases (Section~\ref{sec:time.variability}).

\begin{deluxetable*}{ccccccccc}
\tabletypesize{\footnotesize}
\tablecolumns{9}
\tablewidth{\textwidth}
\tablecaption{Components of the SiS($14-13$) line observed with CARMA \label{tab:components}}
\tablehead{
  &                & \multicolumn{3}{c}{Observations}                 & \multicolumn{4}{c}{Gaussian fit} \\
\colhead{Component} & \colhead{Velocity range} & \colhead{$F_\subscript{int}$} & \colhead{$F_\subscript{max}$}    & \colhead{Peak}       & \colhead{Center}      & \colhead{Size}                    & \colhead{P.A.}      & \colhead{Max. diff.}\\
\colhead{}            & \colhead{(\kms)}         &  \colhead{(Jy~\kms)}        & \colhead{(\jybeam~\kms)} & \colhead{(mas)}      & \colhead{(mas)}       & \colhead{(arcsec$^2$)}            & \colhead{($\degr$)} & \colhead{($\sigma_\subscript{RMS}$)}
}
\startdata
Moment 0     & $[-14.5,14.5]$    & $510\pm 90$       & $159\pm 14$          & $(0,40)$   & $(-22,8)$    & $0.47\times 0.46$     & 121        & 1.7\\
A            & $[-10.5,-1.5]$    & $230\pm 30$       & $80\pm 4$           & $(0,40)$   & $(-32,12)$ & $0.46\times 0.44$& 177        & 2.6\\
B            & $[-1.5,0.5]$    & $87\pm 11$        & $23.1\pm 1.7$        & $(150,40)$ & $(-3,-20)$ & $0.56\times 0.50$ & 117        & 2.9\\
C            & $[0.5,3.5]$    & $70\pm 8$         & $15.8\pm 1.4$        & $(0,-110)$ & $(4,-31)$ & $0.55\times 0.52$ & 118        & 2.3\\
D            & $[3.5,14.5]$    & $161\pm 19$       & $58\pm 3$            & $(-60,40)$ & $(-23,19)$ & $0.47\times 0.43$& 96        & 2.6\\
\enddata
\tablecomments{
The fluxes of each component were calculated by integrating the emission above 20\% of the peak flux over the Doppler velocity ranges in the second column.
These velocities are referred to the systemic velocity, which is $v_\subscript{sys}=-26.5$~\kms{} in the LSR coordinate system \citep*[e.g.,][]{cernicharo_2000}.
The HPBW of the synthetic PSF is $0\farcs25$.
The position of the star is $\alpha(\textnormal{J}2000)=09^\subscript{h}47^\subscript{m}57.\!\!^\subscript{s}435$ and $\delta(\textnormal{J}2000)=+13^\circ16^\prime43\farcs86$, determined by \citet{fonfria_2014} from the continuum emission.
The uncertainty of the peak position is estimated to be $\simeq 30$~mas \citep{fonfria_2014}.
The statistical uncertainties of the center of the Gaussians, their major and minor axes, and their P.A. are $\simeq 15$~mas, 30~mas, and 30$\degr$, respectively.
The last column contains the maximum differences between the Gaussian fits and the observed brightness distributions expressed in terms of the noise RMS of each map, $\sigma_\subscript{RMS}$, i.e., the uncertainty of $F_\subscript{max}$.
}
\end{deluxetable*}

The moment 0 map of the SiS($14-13$) line shows a partially resolved compact brightness distribution roughly centered on the star (Figures~\ref{fig:f2} and \ref{fig:f3}; Table~\ref{tab:components}).
The lower level contours are mostly circular.
There is an emission deficit to the E of the star that is probably caused by a negative sidelobe of the synthetic PSF close to the main beam, smaller than 10\% and arranged along the direction with P.A.~$\simeq 120\degr$.
The higher level contours are elliptical with their major axes oriented roughly along the NE-SW direction and a ratio of the minor to the major axes of $\simeq 80$\%, suggesting that the brightness distribution is elongated along the direction with P.A. ranging from 50 to 70$\degr$.
The emission peaks at $0\farcs07\pm 0\farcs04$ to the NW of the star \citep{fonfria_2014}.

\begin{figure}
\centering
\includegraphics[width=0.4\textwidth]{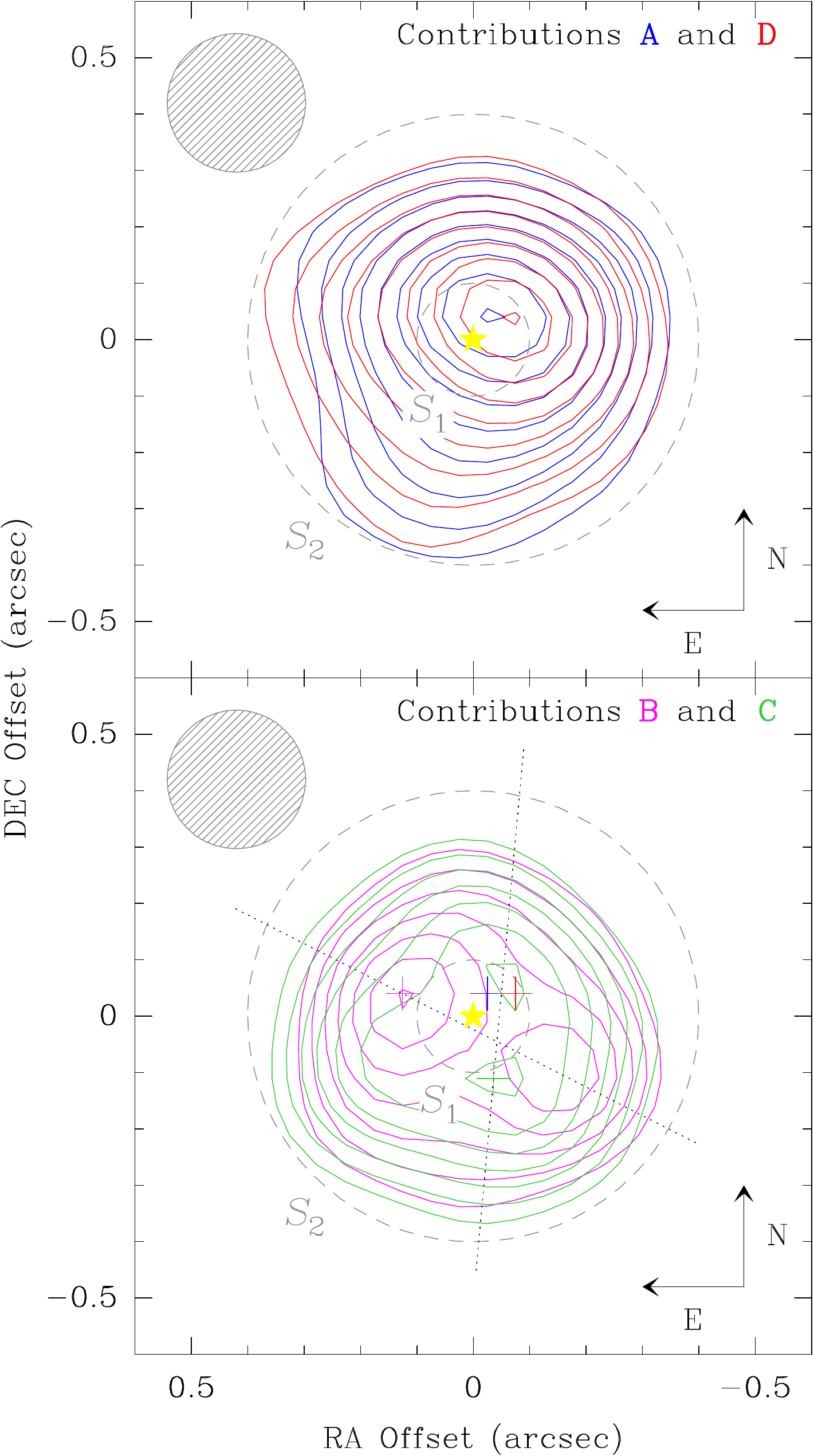}
\caption{CARMA SiS($14-13$) emission from the different maser components integrated over several frequency intervals (see Table~\ref{tab:components} and Section~\ref{sec:results}).
Components A and D are plotted in blue and red in the upper panel whereas components B and C are in magenta and green in the lower panel (see Table~\ref{tab:components}).
The contours for the A and D components are at 20 to 90\% by 10\%, and 99\% their maximum fluxes while for components B and C they are at 40 to 90\% by 10\%, 
and 99\% their maximum fluxes.
The colored crosses indicate the position of the maxima of every component.
The black dotted lines describes possible remarkable directions.
The dashed gray circles $S_1$ and $S_2$ represent the shells at 5 and 20\rstar{} from the star.
The HPBW of the synthetic PSF of the observations is outlined at the upper left corner of each panel.
The yellow star marks the position of the central star.}
\label{fig:f4}
\end{figure}

The position-velocity diagrams show that the SiS maser emission mostly comes from the four relatively isolated components A, B, C, and D presented above (Figures~\ref{fig:f2} and \ref{fig:f3}).
The strongest component, A, with a peak flux of $\simeq 16.9$~\jybeam, is located in front of the star at $-5.5$~\kms{} with respect to the systemic velocity.
The following component in brightness, D, displays a maximum flux of $\simeq 8.5$~\jybeam{} and is located behind the star between 8.5 and 10.5~\kms.
The bulk emission of both components A and D comes from the northern hemisphere of the envelope. 
The resemblance between the position in the plane of the sky and the shape of components A and D (blue-shifted and red-shifted with respect to the central star), suggests that they are similar emitting structures mirrored with respect to the plane of the sky containing the star.

Component B is a weaker structure around the star elongated along the direction with a P.A. between 70 and 90$\degr$, showing an average flux of $\simeq 35$\% of the cube peak emission ($\simeq 16.9$~\jybeam).
Its size is $\simeq 0\farcs5\times 0\farcs2$ in the plane of the sky and of $\simeq 3$~\kms{} along the velocity axis.
The maximum flux of this structure ($\simeq 45$\% of the peak emission) is found at $\simeq 0\farcs1$ to the E of the star, about 3 times the estimated position uncertainty. 
The position-velocity diagrams show the structure corresponding to component W, that is located at $\simeq 3.5$~\kms{} and where the emission is much weaker than at other velocities.
Between component W and the central star is component C, weaker than component B.
The whole structure is embedded in an extended, roughly uniform, and weak structure with an emission $\lesssim 20$\% of the cube peak emission.

The integrated emissions coming from every component A, B, C, and D are mapped in Figure~\ref{fig:f4}.
The position of their emission peaks and their corresponding elliptical Gaussian fits can be found in Table~\ref{tab:components}.
Regarding the morphology of components A and D, since their lowest contour levels are above the $3\sigma$ level, the differences observed in them suggest a slightly different structure of the emitting regions along the E-SE direction.
This difference might evidence an anisotropy of the gas density as long as an elongation roughly along the SE direction is also noticeable in the H$^{13}$CO and SiC$_2$ maps presented by \citet{fonfria_2014}.
Nevertheless, this elongation is not present in the SiO maps, which could imply that chemistry also plays a role in the SE quadrant of the envelope.

Component B is substantially different than components A and D.
The lowest level contours are roughly circular and are centered on the star.
However, component B clearly shows a bipolar structure with two peaks at a distance of $0\farcs13\pm 0\farcs04$ and $0\farcs17\pm 0\farcs04$ from the star ($6.5\pm 2.0$ and $8.5\pm 2.0$\rstar) with P.A.~$\simeq 70$ and $230\degr$, respectively.
The peak along the direction with P.A.~$\simeq 70\degr$ accounts for 65\% of the total emission of this component.
It is worth noting that the emission peak of components A, C and D  lie along the N-S direction, roughly perpendicular to the direction defined by the peaks of component B (NE-SW).

There is a remarkable resemblance between the structure showed by component B and the brightness distribution of the NaCl($21-20$) line in the vibrational ground state in the channels around the systemic velocity presented by \citet{quintana-lacaci_2016} (see their Figure~2).
The most evident difference between both structures is the position of their emission peaks, which are $\simeq 0\farcs3\simeq 15\rstar$ from the star in the NaCl line.
This supports the idea that the gas is anisotropically distributed around the star, as was already suggested by \citet{fonfria_2014} after analysing the high angular resolution emission of H$^{13}$CN, SiO, and SiC$_2$.

Component C is also different than the other components.
It shows an overall circular morphology avoiding the central region and two emission peaks to the N and S.
The southern peak is close to the SW maximum displayed by component B, suggesting that the structure of both components could be somehow related.
Moreover, the southern emission of this component roughly peaks at the mirrored position of the emission peaks of components A and D with respect to the direction defined by the peaks of component B (see Figure~\ref{fig:f4}).
On the other hand, the northern emission peak matches up quite accurately with the emission peaks of components A and D.
This could mean that all these components are related to a common structure in the surroundings of the central star.

\begin{figure}
\centering
\includegraphics[width=0.45\textwidth]{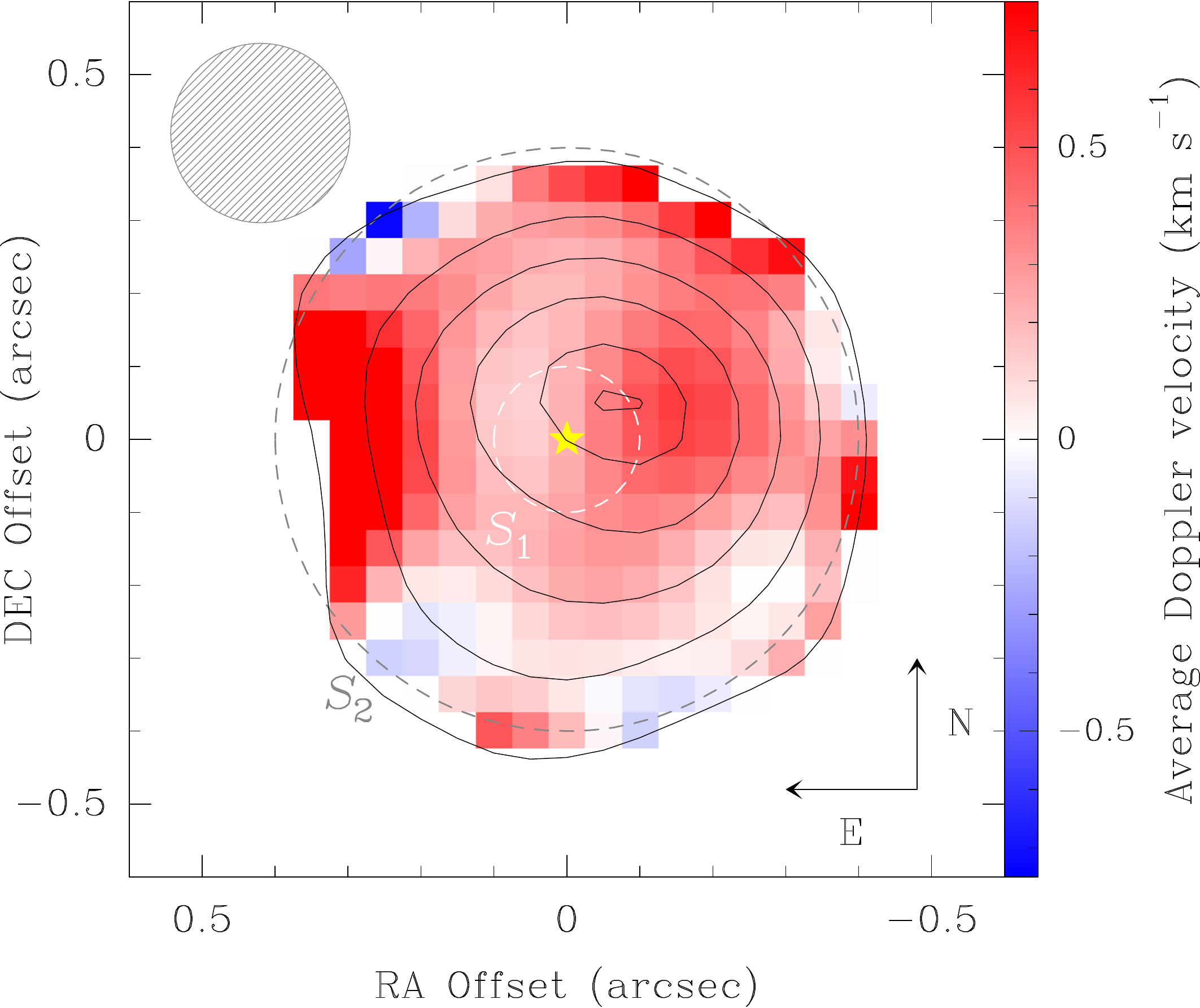}
\caption{Average Doppler velocity with respect to the systemic velocity (moment 1 map) of the observed brightness distribution.
The contours represent the moment 0 map at levels 10, 30, 50, 70, 90, and 99\% of the peak emission.
Only the emission above the 10\% contour has been taken into account.
The systemic velocity is $-26.5$~\kms.}
\label{fig:f5}
\end{figure}

Figure~\ref{fig:f5} shows the average Doppler velocity of the observed brightness distribution (moment 1 map) with variations not larger than $\simeq 1.5$~\kms.
It is mostly red-shifted with respect to the systemic velocity. 
The average emission that comes from the NW, which coincides with the emission peak of components A and D, is slightly red-shifted with respect to that coming from the South.
This means that the emission is roughly symmetric regarding the plane that contains the central star but the emitting regions behind the star move faster related to Earth than those in front of it.
This is particularly evident to the NW of the star, where the emission of component D is weaker than that of component A (Table~\ref{tab:components}).

\subsection{Time variability of thermal and non-thermal SiS and $^{29}$SiS emission}
\label{sec:time.variability}

\begin{turnpage}
\begin{figure*}
\centering
\includegraphics[height=0.9\textwidth]{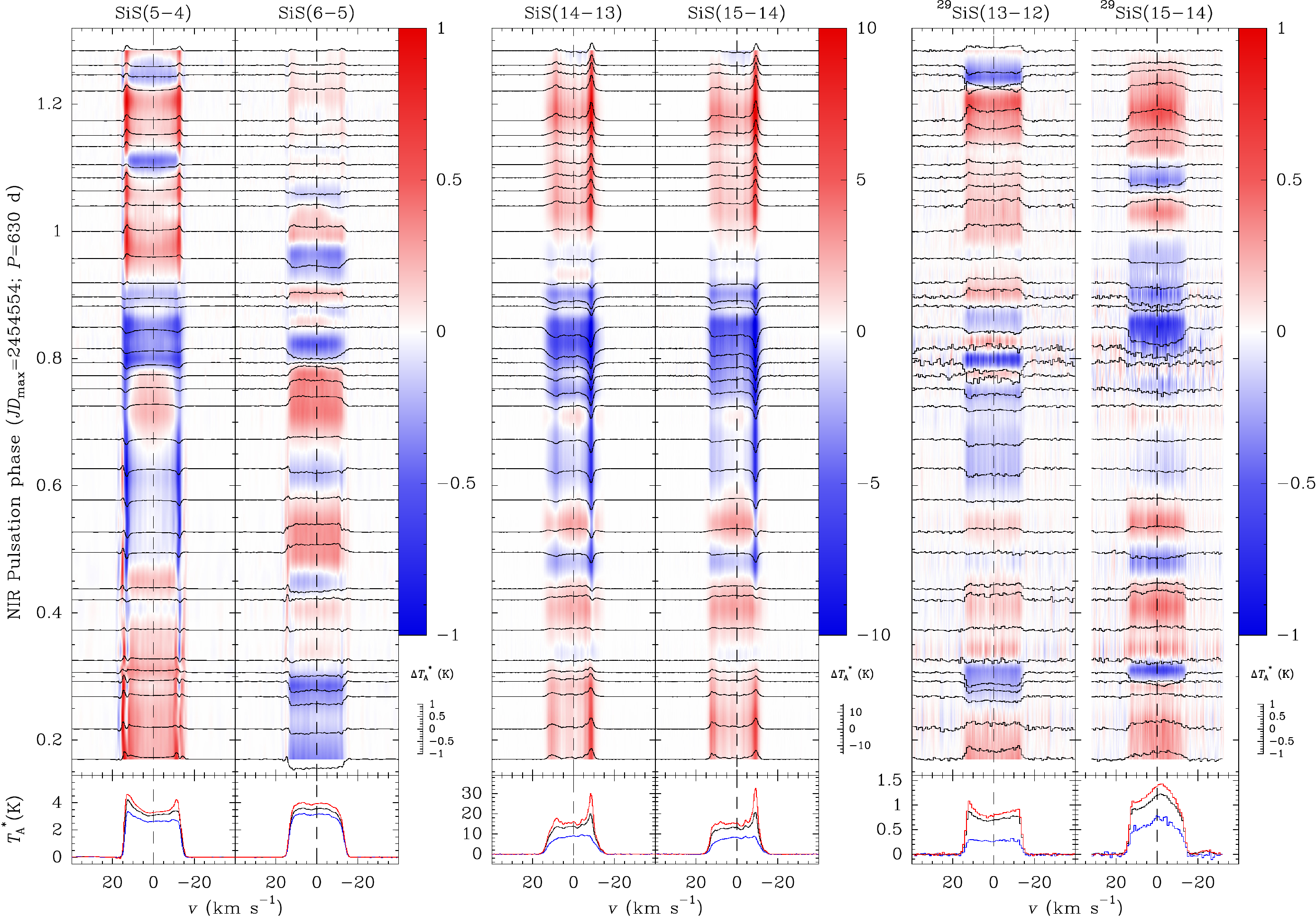}
\caption{Time variability of the SiS lines $J=5-4$, $6-5$, $14-13$, and $15-14$, and of the $^{29}$SiS lines $J=13-12$ and $15-14$.
  The vertical axes of the upper and lower panels are the NIR pulsation phase as view by \citet{menten_2012} and the antenna temperature of the observations.
  The NIR pulsation phase is $(JD-JD_\subscript{max})/P$, where $JD$ is the Julian Date, $JD_\subscript{max}$ is a reference JD for the maximum emission, and $P$ is the pulsation period in days.
  The horizontal axis is the Doppler velocity related to the systemic velocity.
  In the lower panels we have plotted the average, strongest, and weakest lines (black, red, and blue).
  The upper panels contain the absolute difference of every line with respect to the average.
  The color scale represents the interpolation of the observed lines for every pulsation phase performed with the GILDAS routine \texttt{random\_map}.
  The Doppler velocity is referred to the systemic velocity ($\simeq -26.5$~\kms).
}
\label{fig:f6}
\end{figure*}
\end{turnpage}

Figure~\ref{fig:f6} shows the observations of the SiS $v=0$ $J=5-4$, $6-5$, $14-13$, and $15-14$, and $^{29}$SiS $v=0$ $J=13-12$ and $15-14$ lines mentioned above spanning roughly through a single pulsation period.
All the data have been arranged in six columns, one per molecular line.
In a given column, we have plotted the difference of the corresponding line observed at every pulsation phase with respect to the average line over the whole monitoring.

The $^{29}$SiS lines show what seems to be a long term time variation.
There are also departures from the long term behavior that suddenly affect the lines observed at isolated pulsation phases or during a few adjacent phases, e.g., in the $^{29}$SiS lines at phase 0.30 or in the SiS($5-4$) at phases 1.10 and 1.25.
Since these observations have been corrected from calibration issues (see Appendix~\ref{sec:cal.issues}), all of these short term departures are in principle real.

The $J=5-4$ and $6-5$ SiS lines show a roughly complementary dependence with time for pulsation phases below 0.8.
Above this pulsation phase, however, it is not clear if their time dependences are or not related.
The long term variation of the SiS($5-4$) line seems to show the same period than the NIR light curve and a phase difference of about 0.2 (see Section~\ref{sec:discussion.variability}), although it displays short time scale departures from it mostly in the central part of the line.
It is worth to keep in mind that these lines were not corrected from pointing or calibration errors, as we explain in Appendix~\ref{sec:cal.issues}.
It is noticeable the presence of narrow peaks at high expansion velocities in the SiS($5-4$) line and their absence in the SiS($6-5$) line throughout the whole monitoring.
In addition, these peaks vary strongly with the pulsation phase, much more than the rest of the line, contrarily to what happens with the $J=6-5$ line, for which the line profile is approximately constant.
These facts suggest that the level $J=5$ varies with the pulsation phase in a different way than the levels $J=4$ and 6, as it was already suggested by \citet{carlstrom_1990}, who claim this difference as due to a possible IR overlap between a SiS line involving the level $J=5$ and a line of other molecule.

The profile of the SiS($6-5$) line resembles an inverted U, a shape typically associated to optically thick lines coming from spatially resolved regions \citep*[e.g.,][]{zuckerman_1987}.
However, the SiS($5-4$) displays a double-horned profile.
Considering that they are formed roughly in the same region of the envelope and the main beam of the 30~m telescope is nearly the same for their frequencies (HPBW~$\simeq 25\arcsec$), we suggest that the high-velocity SiS($5-4$) peaks are masers in nature, as the peaks observed in the SiS lines $J=14-13$, $15-14$, and $1-0$ (see below and in \citealt{henkel_1983} and \citealt{gong_2017}).

\begin{figure}
\centering
\includegraphics[width=0.475\textwidth]{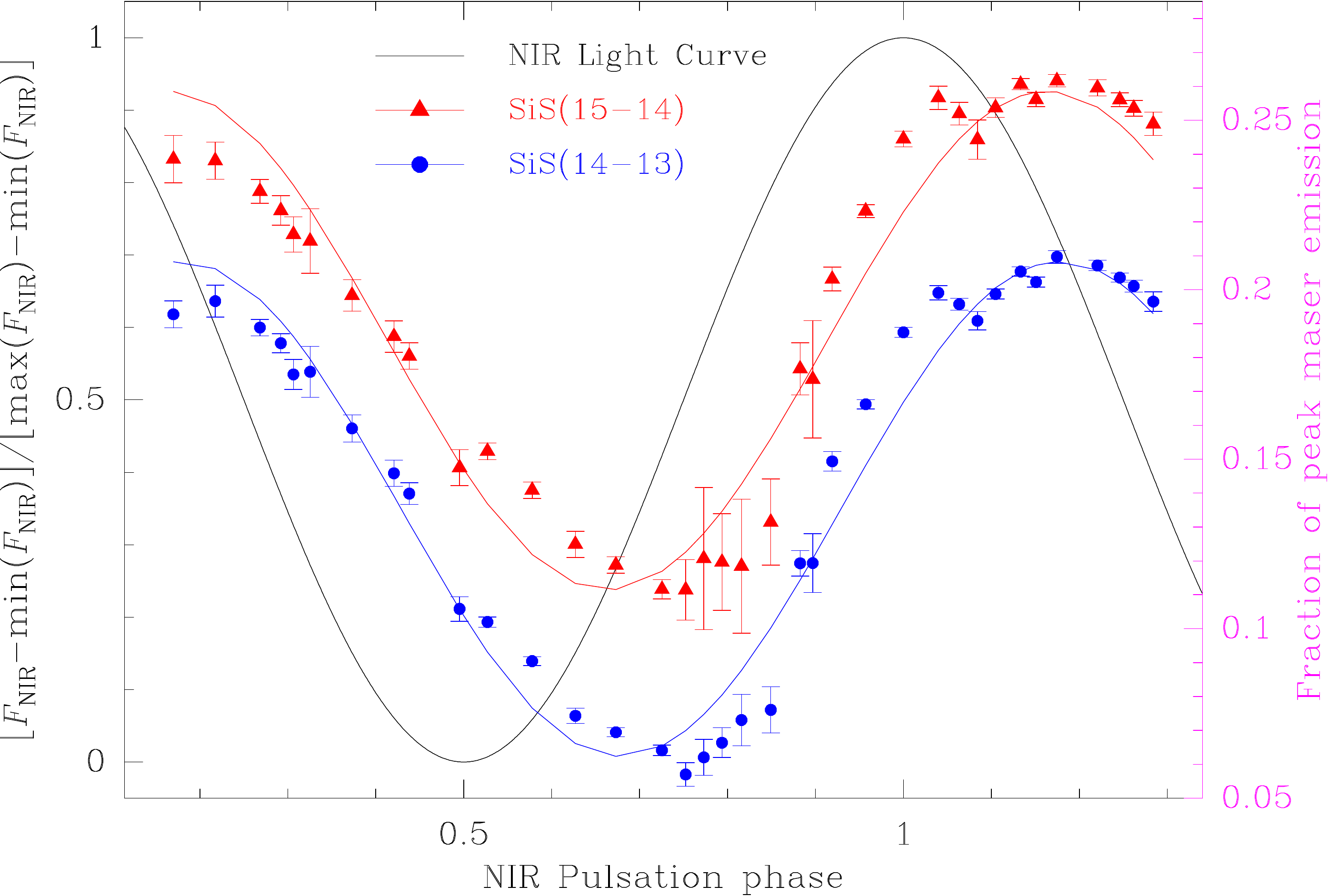}
\caption{Time dependence of the relative flux of the peak labeled as component A$^*$ with respect to the total emission measured in the same velocity range in the $J=14-13$ and $15-14$ SiS lines along one pulsation period.
    The blue dots and the red triangles are the data for the $J=14-13$ and $15-14$ SiS lines.
    The solid lines are the fits to these data sets.
  The solid black curve is the normalized NIR luminosity of \irc{} \citep{menten_2012}.}
\label{fig:f7}
\end{figure}

The strongest time variation is shown by the maser peaks of the $J=14-13$ and $15-14$ SiS lines (labeled as A$^*$ and D$^*$ in Figure~\ref{fig:f1}).
The emission of both lines seem to be phase coherent and also coherent with that shown by the SiS($5-4$) line following the NIR light curve during most of the monitoring, although they show a few short time scale departures from the long term dependence.
As it occurs with the SiS($5-4$) line, the long-term variation of the SiS lines $J=14-13$ and $15-14$ shows also a phase difference of $\simeq 0.2$ with respect to the NIR light-curve.

Figure~\ref{fig:f7} shows the relative flux of the narrow peak which is the component A$^*$ with respect to the total emission measured in the same velocity range in the SiS lines $J=14-13$ and $15-14$ against the NIR pulsation curve.
The flux below the feature was calculated taking advantage of its small width, which let us determine the shape of the line related to the extended emission.
The very good time sampling of a pulsation period achieved during our monitoring allowed us to describe accurately the variation undergone by this component.
This variation is periodical, as it occurs with the NIR luminosity.  
Nevertheless, as we noted above, the SiS and NIR curves are not in phase.
We have fitted the data sets with offset sines finding phase differences of $0.179\pm 0.006$ and $0.164\pm 0.007$ for the lines $J=14-13$ and $15-14$ with respect to the NIR light-curve.
Moreover, the maser light-curves cannot be described accurately with a simple trigonometric function: the slope of the dependence after the minimum is steeper than during the decreasing intensity phase.
We will discuss about the origins of this phase difference in Section~\ref{sec:discussion.variability}.

\section{Modeling}
\label{sec:modeling}

In the current work, we have modeled in detail only the SiS($14-13$) line.
The lower spatial resolution of the SiS($15-14$) line is insufficient to resolve, even partially, the complex brightness distribution found in the $\simeq 1\arcsec$-sized region around the central star.
In fact, this distribution (not shown) is very similar to that of the SiS($14-13$) line (Figures~\ref{fig:f2} and \ref{fig:f3}) but the emission peak in every velocity channel is not offset with respect to the position of the star or the offsets are not significant.
Thus, little spatial information can be derived from the SiS($15-14$) maps to enrich the analysis of the SiS($14-13$) line.

The model of the SiS($14-13$) line presented in Section~\ref{sec:observations} was achieved with an improved version of the code initially developed by \citet{fonfria_2008} to calculate the emission of a spherically symmetric circumstellar envelope composed of dust and radially expanding gas \citep{fonfria_2011,fonfria_2015,fonfria_2017}.
The new version allows us to deal with 3D asymmetric envelopes.
It has been successfully used by \citet{fonfria_2014} to model several lines observed with low spectral resolution of H$^{13}$CN, SiO, and SiC$_2$, in addition to the same SiS($14-13$) line that we analyze in the current paper.
This code was tested comparing the synthetic emission with analytical and numerical results of simple scenarios and with the results of the non-local, non-LTE code by \citet{daniel_2008}.
Our code determines the population of the levels involved in the calculations from the input rotational and vibrational temperatures at selected points of the envelope, so that the statistical equilibrium equations are not solved.
Then, it calculates the synthetic emission solving numerically the radiation transfer equation.
Further considerations on how our code works can be found in Appendix~\ref{sec:considerations}.

\begin{deluxetable}{lccc}
\tabletypesize{\footnotesize}
\tablecolumns{4}
\tablewidth{0.475\textwidth}
\tablecaption{Fixed parameters in model\label{tab:fixed.parameters}}
\tablehead{\colhead{Parameter} & \colhead{Value} & \colhead{Units} & \colhead{Ref.}}
\startdata
$D$ & 123 & pc & 1\\
$\dot M$ & $2.0\times 10^{-5}$ & M$_\odot$~yr$^{-1}$ & 3\\
$T_\star$ & 2330 & K & 4\\
$\alpha_\star$ & 0.02 & arcsec & 4\\
$\rstar$ & $3.7\times 10^{13}$ & cm & \\
$\rin$ & 5 & $\rstar$ & 5\\
$\rout$ & 20 & $\rstar$ & 5\\
$v_\textnormal{\tiny exp}(1\rstar\le r<\rin)$ & $1+2.5(r/\rstar-1)$ & \kms & 1\\
$v_\textnormal{\tiny exp}(\rin\le r<\rout)$ & 11 & \kms & 1\\
$v_\textnormal{\tiny exp}(r\ge\rout)$ & 14.5 & \kms & 1\\
$\Delta v(1\rstar\le r<\rin)$ & $5(\rstar/r)$ & \kms & 3\\
$\Delta v(r\ge\rin)$ & 1 & \kms & 3\\
$\tk(1\rstar\le r< 9\rstar)$        & $T_\star(\rstar/r)^{0.58}$  & K & 6   \\
$\tk(9\rstar\le r< 65\rstar)$       & $\tk(9\rstar)(9\rstar/r)^{0.40}$ & K & 6   \\
$\tk(r\ge 65\rstar)$               & $\tk(65\rstar)(65\rstar/r)^{1.2}$ & K & 6   \\
$\trot\left(1\rstar\le r<\rin\right)$ & $2330\left(\rstar/r\right)^{0.62}$ & K & 1 \\
$\trot\left(r\ge\rin\right)$ & $860\left(\rin/r\right)^{0.55}$ & K & 1 \\
$\tau_\textnormal{\tiny dust}(257~\textnormal{GHz})$ & $1.5\times 10^{-3}$ &  & 2\\
$T_\subscript{dust}\left(r\ge\rin\right)$ & $825\left(\rin/r\right)^{0.39}$ & K & 1\\
\enddata
\tablecomments{
$D$: distance to the star; 
$\dot M$: mass-loss rate; 
$T_\star$: effective stellar temperature; 
$\alpha_\star$: angular stellar radius; 
$\rstar$: linear stellar radius;
$\rin$ and $\rout$: boundaries between regions with a different
expansion velocity (see text);
$v_\textnormal{\tiny exp}$: expansion velocity field;
$\Delta v$: line width;
\trot: rotational temperature;
$\tau_\textnormal{\tiny dust}$: optical depth of dust;
$T_\subscript{dust}$: temperature of the dust grains.
(1) \citet{fonfria_2015}
(2) \citet{fonfria_2014}
(3) \citet{agundez_2012}
(4) \citet{ridgway_1988}
(5) \citet{fonfria_2008}
(6) \citet{debeck_2012}
}
\end{deluxetable}

The envelope model assumed in this work is that derived by \citet{fonfria_2008}, recently improved by \citet{fonfria_2015}.
The envelope is divided into three Regions (I, II, and III, outwards from the star) ranging from the stellar photosphere to 5\rstar, from 5 to 20\rstar, and beyond 20\rstar, respectively (see Table~\ref{tab:fixed.parameters}).
The adopted expansion velocity field is spherically symmetric and it is composed of a linearly growing velocity ($1+2.5(r/\rstar-1)$~\kms) in Region I, and two constant expansion velocities of 11.0 and 14.5~\kms{} in Regions II and III.
We have assumed a line width due to turbulence and the gas kinetic temperature which adopts a value of 5~\kms{} at the stellar surface and decays as $\propto 1/r$ up to 1~\kms{} at 5\rstar, remaining constant outwards \citep{agundez_2012}.
The rotational and vibrational temperatures have been assumed to follow a power-law $\propto r^{-\alpha}$, where $\alpha$ changes in each region of the envelope.
All these physical quantities have been assumed as fixed during the search for the best model except for the rotational temperature, that has been modified at certain points or small regions of the envelope to invert the population of the levels involved in the observed line with the aim to reproduce the maser emission.

We did not modify the gas density or the SiS abundance in these small maser emitting regions because there is no reliable information about how these quantities can vary in the dust formation zone.
The observations with the highest angular resolution acquired to date of the innermost envelope of \irc{} are those presented by \citet{fonfria_2014}, taken under the same project than the CARMA observation we are presenting in the current paper.
In that paper, we found variations of the abundance with respect to H$_2$ of H$^{13}$CN, SiO, and SiC$_2$ after assuming the gas density to follow the law $\propto r^{-2}v_\subscript{exp}$, where $v_\subscript{exp}$ is the gas expansion velocity field.
However, we could not explore the thermal SiS emission and no SiS abundance distribution has been derived so far, different from the spherically symmetric one \citep{fonfria_2015}.
Therefore, introducing departures from the spherically symmetric density and abundance distributions to reproduce the maser emission would be highly speculative.
Moreover, even varying these quantities, the excitation temperature for the masing line would need to be input as well, making the optical depth to depend on two free parameters.
Thus, we adopt the typical spherically symmetric gas density and SiS abundance with respect to H$_2$ profile.
We controlled the optical depth of the maser emitting regions only with the rotational temperature.
Under these circumstances, we obtained a good estimate of the optical depth in the masing regions, but the derived rotational temperature could be inaccurate in them.
The goal of this procedure is to reproduce the observations using the smallest number possible of maser components.

The rest frequencies of the observed lines ($J=14-13$ and $15-14$), which are 254.103213 and 272.243055~GHz, and the spectroscopic constants of SiS were taken from the MADEX code \citep{cernicharo_2012,velilla_2015}.
The dust has been assumed to be composed of amorphous carbon.
The dust optical refractive indexes needed to calculate the dust emission at the observed frequencies were estimated by linear extrapolation from the results by \citet{suh_2000} at wavelengths up to 1~mm \citep{fonfria_2014}.

\subsection{The best model of the SiS($14-13$) line}

\begin{figure*}
  \centering
\includegraphics[width=0.9\textwidth]{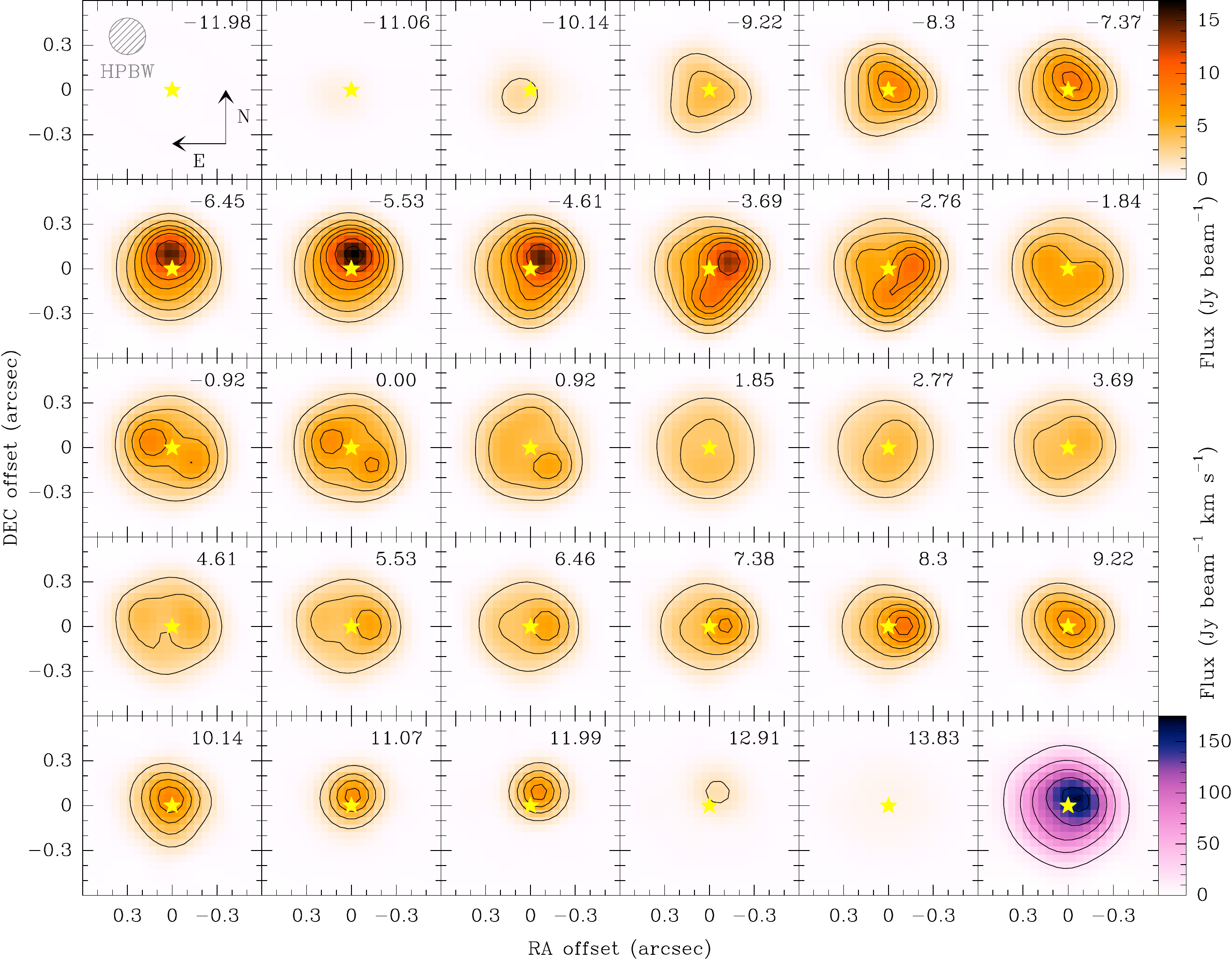}
\caption{Velocity-channel maps of the \textit{modeled} brightness distribution.
The brightness distribution has been modeled as if it would be observed with CARMA in its B configuration.
See caption of Figures~\ref{fig:f2} and \ref{fig:f3}, and Section~\ref{sec:observations} for more information.}
\label{fig:f8}
\end{figure*}

\begin{figure*}
  \centering
\includegraphics[width=0.9\textwidth]{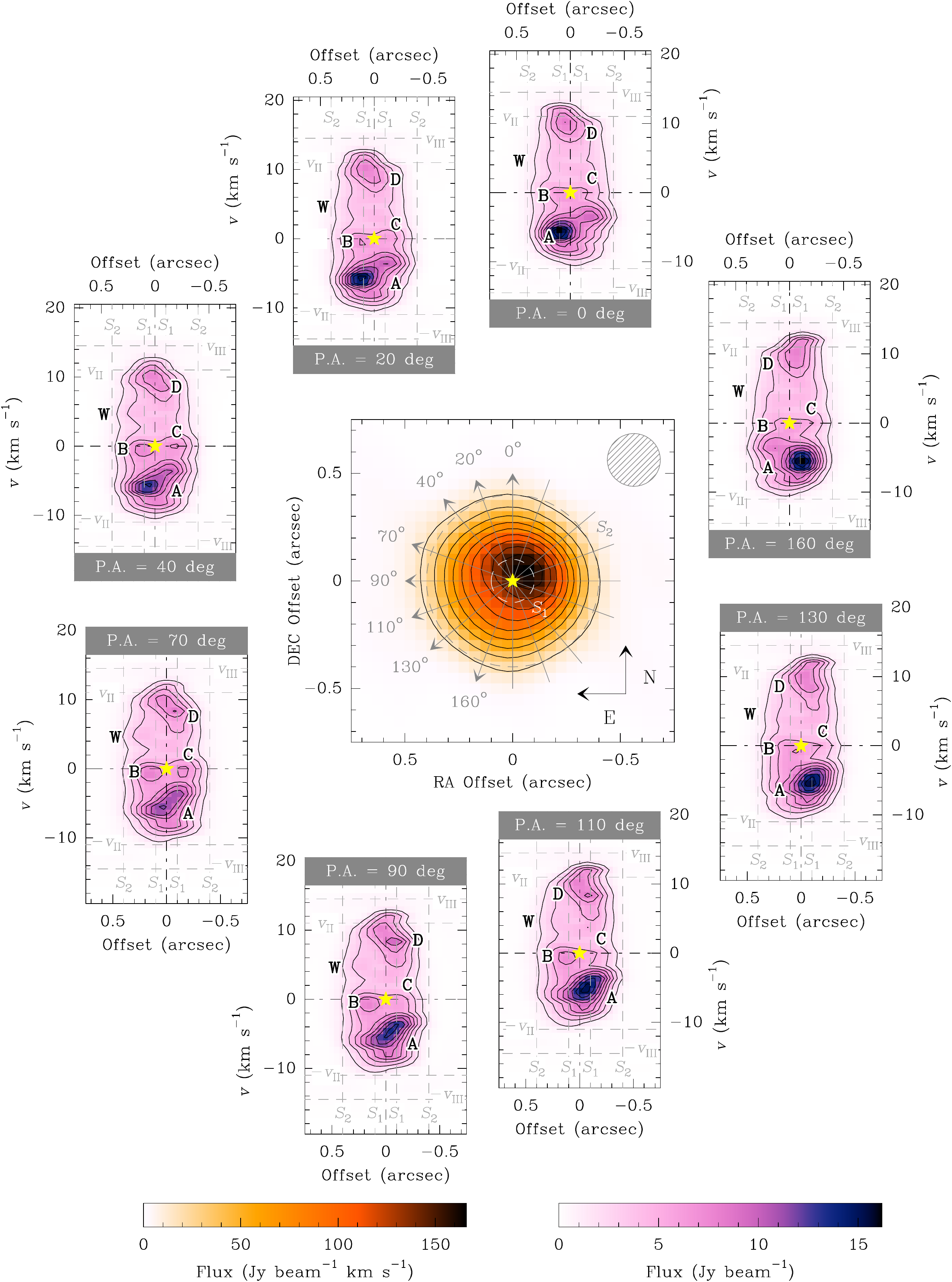}
\caption{Moment 0 map of the \textit{modeled} brightness distribution of the SiS($14-13$) line (central panel) and position-velocity maps for the slices with P.A.~=~0, 20, 40, 70, 90, 110, 130, and 160$\degr$.
The brightness distribution has been modeled as if it would be observed with CARMA in its B configuration.
See caption of Figures~\ref{fig:f2} and \ref{fig:f3}, and Section~\ref{sec:observations} for more information.}
\label{fig:f9}
\end{figure*}

\begin{figure*}
\centering
\includegraphics[width=\textwidth]{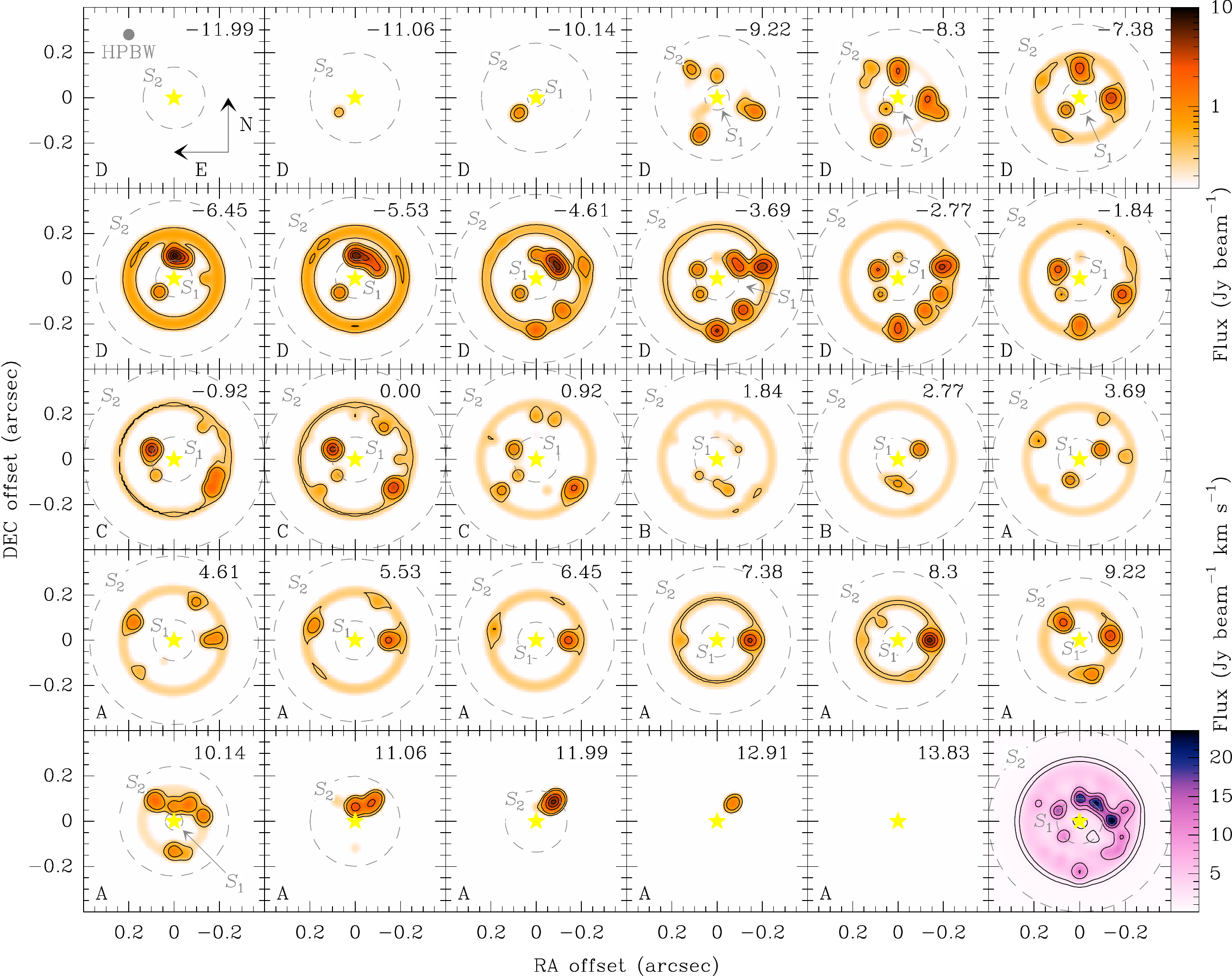}
\caption{
Velocity-channel maps of the synthetic brightness distribution assuming a HPBW~$=0\farcs05$ (orange scale).
The distribution of the integrated emission (moment 0 map) is plotted in the bottom right panel (magenta scale).
The contours are at levels 10, 30, 50, 70, 90, and 99\% of the peak emission of the whole cube (velocity-channel maps) and of the total emission (moment 0 map).
Note that we have used a logarithmic color scale in the velocity-channel maps to improve the visibility of the faintest structures.
The boundaries of Regions I and II ($S_1$ and $S_2$) are marked by gray circles.
These boundaries show different radii in every channel because they are actually spherical shells.
The Doppler velocity expressed in \kms{} of each map is close to the top right corner and is referred to the systemic velocity.
Each velocity-channel belongs to the maser component noted in their bottom left corner.
The spatial scale of the velocity-channel maps is different to that of Fig.~\ref{fig:f8} on purpose to favor the exploration of the derived emission structure.}
\label{fig:f10}
\end{figure*}

The observed emission cannot be modeled assuming only positive excitation temperatures.
The synthetic emission is far from reproducing the observed one even after assuming excitation temperatures as high as $10^5$~K.
The observational uncertainties cannot explain this difference.
This fact forced us to consider maser emitting components in different places of the envelope model that are included only if it is absolutely necessary.
Following this strategy, the most important features of the observed brightness distribution were reasonably well reproduced (Figures~\ref{fig:f8}, \ref{fig:f9} and \ref{fig:f10}).
Most of these components are masing clumps elongated along the line-of-sight that span through different envelope shells.
The size of these clumps in the plane of the sky derived from our model is of $\simeq 1.5\rstar$ in average.
There is also an arc-like masing structure that could be thought to comprise several very close clumps.
Although the derived size distribution for the clumps seems to reproduce reasonably well the observations, we note here that higher angular resolution may provide more accurate results.
In any case, the actual compact emitting regions are small, which is in good agreement with the morphology of the typical maser emitting regions found in other AGB stars \citep*[e.g.,][]{soria-ruiz_2007}.
In addition to the set of clumps, we need an extended maser structure to reproduce the observed extended emission (see below).
The high number of different channels to model, each showing complex emission well above the RMS noise level, explains the high number of clumps needed to reproduce the observations.

The best way we found to model the extended emission deduced to exist from the observation includes a maser emitting shell with a radius of 13\rstar{} and a thickness of $\simeq 2\rstar$, i.e., showing an approximate size of $30\rstar\simeq 0\farcs6$.
It is not possible to estimate the actual thickness from our data because of the lack of angular resolution (see Section~\ref{sec:discussion} for a discussion about this).
We do not expect a bigger structure because, in this case, we would have a contribution to the emission at larger distances from the star nonexistent in our observations.
We modified the rotational temperature at any position in this shell when needed to reproduce the observation.
In fact, the modeling procedure resulted in a maser emitting shell that is stronger in the velocity channels around $8.0$, $-0.5$, and $-5.0$~\kms, and weaker at $11.0$, $3.5$ (component W), and $-9.5$~\kms.

Component A cannot be reproduced with a single clump located at the emission peak because it results in a significantly narrower component than observed.
The best fit suggests that component A is mainly described by one arc spanning different velocity channels to the NW of the star (see Figure~\ref{fig:f10} in the range $[-6.9,-3.2]$~\kms), which accounts for the emission peak and its close environment.
This arc extends spatially from $\simeq 6.5$ to 11\rstar{} from the star and its P.A. varies from $\simeq 300$ to 360$\degr$.
Its emission is more blue-shifted to the N of the star than towards the NW.
Additional weaker emitting clumps are necessary to reproduce the rest of the A emission in the velocity range $[-10.5,-1.5]$~\kms.
These clumps are placed at different locations of the envelope between 5 and 20\rstar, i.e., in Region II.
In particular, the clumps which reproduce the emission at $-1.5$ and $-10.5$~\kms{} are very close to the boundaries of this region.
Most of them span up to 4 velocity-channels although one of them located at the SE of the star, interestingly spans out 11 channels (i.e., $\simeq 10$~\kms), taking part in component B as well.
However, with such a large velocity span the growth of the emission of this clump along the line-of-sight cannot be exponential because in this case the emitted flux would be huge, contrarily to what it is really observed.
This means that this structure is divided into several shorter clumps radiatively disconnected.
This fact suggests that the coherent velocity is several times lower than 10~\kms, as expected to occur for the expanding gas close to the star, with a turbulent velocity $\lesssim 5$~\kms{} \citep{monnier_2000b,agundez_2012,fonfria_2015,fonfria_2017}.

In Table~\ref{tab:fractions} we have included the fraction of the total flux of each emitting structure accounted by every component (thermal emission, extended maser emission, and clumps/arcs).
Despite the brightness density of the arc and some clumps in component A is quite strong compared to the thermal and extended maser emissions, each emitting structure shows comparable fractions of the total emission due to the different solid angles subtended.
As a result of this, most of the total maser emission comes from the extended maser emission rather than from the clumps and the arc.
On the other hand, the maser emission of component A accounts for more than 80\% of its total emission (thermal+maser).

\begin{deluxetable}{cccc}
\tabletypesize{\footnotesize}
\tablecolumns{4}
\tablewidth{0.475\textwidth}
\tablecaption{Fraction of the total emission per component\label{tab:fractions}}
\tablehead{
  \colhead{Component} &  \colhead{Thermal} & \colhead{Extended} & \colhead{Clumps+Arcs}\\
  \colhead{} &  \colhead{} & \colhead{maser emission} & \colhead{}\\
  \colhead{} & \colhead{\%} & \colhead{\%} & \colhead{\%}}
\startdata
Moment 0 & 25.2          & 44.1          & 30.7  \\
A        & \textit{16.4} & 46.6          & \textbf{37.0} \\
B        & 23.3          & 42.5          & 34.2 \\
C        & 34.6          & \textbf{48.2} & \textit{17.2} \\
D        & \textbf{36.0} & \textit{28.2} & 35.8 \\
\enddata
\tablecomments{These fractions are related to the integrated fluxes contained in Table~\ref{tab:components}.
The minimum value of every structure (thermal, extended maser emission, and clumps+Arcs) comparing its contribution to all the emission components is highlighted in italics and the maximum value in boldface.}
\end{deluxetable}

Component B is mostly composed by two clumps spanning $\simeq 5$~\kms{} and located at 5.7 and 11\rstar{} from the star along the directions with P.A.~$\simeq 65$ and 235$\degr$, respectively.
The total flux distribution per component is similar than for component A, although the fraction of the thermal emission is higher.

Component C shows a very simple structure comprising two weak clumps located at 5.7\rstar{} with P.A.~$\simeq 180$ and 295$\degr$.
Most of its maser emission comes from the extended maser emission structure.
This component comprises the part of the brightness distribution in which the clumps account for the smallest fraction of the total emission.

Regarding component D, it can be explained with a simpler structure than component A.
It also comprises different clumps distributed across Region II but no arc is needed to explain the red-shifted emission.
The strongest clumps are located to the W and NW of the star.
The clump to the W is particularly interesting because it spans 8 velocity channels ($\simeq 7.4$~\kms) although the exponential growth along the line-of-sight is also disabled, as in component A.
There is a group of clumps that resembles an arc at $10.1$~\kms{} in the same position in the plane of the sky than the arc of component A.
The fraction of maser emission accounted by this maser structure is $\simeq 35$\%, similar to that for components A and B.
However, contrarily to the rest of the components, the extended maser emission is less important and less than a third of the total emission is thermal.

Putting all these data together, 25\% of the total SiS($14-13$) line emission coming from the envelope covered by our CARMA observation has a thermal origin.
The remaining 75\% is maser emission: 60\% comes from the extended maser emission and 40\% from the compact structures.
The low fraction of thermal emission is directly related to the filtering of large scales performed by the interferometer whereas the emission from compact structures are fully recovered.
Thus, most of the emission of this line observed with single-dish telescopes is thermal.

The optical depth of every contribution to the synthetic emission depends on the physical conditions prevailing throughout the envelope and on the solid angle subtended by them, which is not known in our case.
Hence, even though the model reproduces the observations reasonably well (see Figs.~\ref{fig:f2}, \ref{fig:f3}, \ref{fig:f8} and \ref{fig:f9}, we cannot provide meaningful sizes and opacities for the clumps. 
However, in our model, the optical depth adopts different values varying from nearly 0 to a maximum of $\simeq 27.8$ very close to the star for the thermally excited
component of the envelope.
In the case of the maser emitting regions, the optical depth ranges from negative values close to zero, related to very negative excitation temperatures (several times $-100$~K) 
or low population inversions, to $\simeq -11.3$ at about 7\rstar{} from the star, which is associated to gas volumes with strongly inverted populations or low negative excitation temperatures (only a few times $-10$~K).

Modeling the most red-shifted channels required an increment of the width of the outer acceleration shell at $\simeq 20\rstar$.
Initially, we adopted a shell width of $\simeq 0.5\rstar$ but the observation suggest that its width is $\simeq 2.5\rstar$.
This modified acceleration shell has a very limited effect on models derived from previous single-dish observations so it could be easily overlooked \citep*[e.g.,][]{fonfria_2008,fonfria_2015,fonfria_2017,agundez_2012}.

\section{Discussion}
\label{sec:discussion}

\subsection{Caveats about the model imposed by the observations}

The interferometer data sets taken with CARMA and ALMA were not complemented neither with on-the-fly maps acquired with single-dish telescopes nor with total power observations.
The lack of short baseline visibilities prevented us from properly covering the central region of the $uv$ plane, filtering a significant amount of flux. 
The consequence of this flux loss is two-fold: 
(1) the observed absolute flux is a lower limit to the total absolute flux and it is not possible to compare it with the flux determined from lower angular resolution or single-dish observations (e.g., our IRAM 30~m data), and
(2) the outer parts of the observed emission distribution (corresponding in our case to the lowest level contours) could be affected by significant sidelobes in the PSF.
However, we can properly deal with these effects by applying the same constraints to the models than the CARMA array applied on the actual brightness distribution, filtering the same flux than during the observations \citep*[for a deeper explanation, see][]{fonfria_2014}.
This results in realistic predictions not affected by observational issues.

The SiS($14-13$) data presented in this work were observed at the same time with a lower spectral resolution SiS($14-13$) data set
\citep*[$\simeq 12.3$~\kms;][]{fonfria_2014}.
A model composed of several compact emitting regions was proposed to explain the low spectral resolution emission.
The lack of spectral resolution affecting these observations hampered our effort to get accurate positions of the maser emitting regions and their physical properties.
We can now compare models from both high and low spectral resolution observations.
This comparison suggests that the main emitting components derived in the current paper were already proposed to exist by \citet{fonfria_2014}.
Our current model significantly refines the position and size of the components of the maser emitting structure and the gas excitation conditions providing us, in addition, with information about the spatial distribution along the line-of-sight, something impossible to determine from the low spectral resolution data set.

\subsection{The maser pumping mechanism}

Since a significant part of the SiS($14-13$) emission has maser nature, one important question is how the pumping mechanism works in \irc.  
\citet{fonfria_2006} suggested that the overlap between lines of SiS and the abundant species C$_2$H$_2$ or HCN in the MIR could produce the population inversion involved in the maser emission.
In particular, they proposed that the SiS ro-vibrational line $1-0$R$(13)$ could be blended under certain conditions with the C$_2$H$_2$ line $\nu_5$R$_e(9)$, with rest frequency 5.9~\kms{} blue-shifted with respect to the SiS line frequency.
This blending would pump SiS molecules initially in the rotational level $J=13$ of the vibrational ground state to the rotational level $J=14$ in the vibrational state $v=1$.
The excited SiS molecules would deexcite again to the vibrational ground state populating the rotational level $J=15$, since the SiS ro-vibrational transitions must fulfill the selection rule $\Delta J=\pm 1$.
This mechanism would invert the populations of levels $J=13$ and $15$ explaining the maser emission found in the SiS pure rotational lines $J=15-14$ and $14-13$.
The small frequency shift between the SiS and the C$_2$H$_2$ ro-vibrational lines can be shortened by means of the Doppler effect if the C$_2$H$_2$ line is produced in a different place of the envelope than the SiS one.
This indeed occurs in a radially expanding envelope, since every gas volume moves away from any other.

\subsection{Maser structures}

\subsubsection{Extended Maser Emission}

It can be argued that the extended emission is actually thermal emission.
The regions of the velocity channel maps in Figures~\ref{fig:f2} and \ref{fig:f3} delimited by the two lowest level contours, that could be mostly assigned to the observed extended emission, display an average flux of $\simeq 2.6$~\jybeam. 
This flux cannot be reproduced adopting the physical and chemical conditions proposed by \citet{fonfria_2015}, which were derived from a large number of ro-vibrational SiS
lines in the MIR spectral range.
In fact, the assumption of these conditions results in a thermal flux of about 0.7~\jybeam{} at a distance from the star of $\simeq 0\farcs3\simeq 15\rstar$, i.e., $\simeq 75$\% smaller than observed.
These observations were flux calibrated following the standard method and at the same time that the low spectral resolution observations presented by \citet{fonfria_2014}.
The maps of the H$^{13}$CN, SiO, and SiC$_2$ lines in that work do not show similar emission excesses and they can be explained assuming thermal emission.
However, the low spectral resolution SiS line presented in that work was acquired with a different spectral window than that of the high spectral resolution line and it also shows a extended emission that cannot be explained by thermal emission.
Furthermore, the effect of the convolution of the maser clumps with the PSF on the extended emission cannot explain it, either.
Thus, this extended emission seems to be real and consequence of extended maser emission surrounding the central star.

The SiS($14-13$) emission model proposed in this work includes an empty thin shell that accounts for the extended maser emission taking advantage of the limited angular resolution of our observations. 
This structure let us easily control the emission of the synthetic extended emission along every direction.
However, assuming the line overlapping explained before acts as the pumping mechanism in \irc, large portions of its envelope may have proper conditions for the line amplification because of the expected spherical symmetry of the gas expansion velocity field.
Therefore, this thin shell is probably a simple alternative to represent a thicker maser emission region with a similar size.
The size of these structures would be naturally constrained due to the effect on the maser amplification process of the decrease with the distance to the star of the gas density, typical of a geometrical expansion, and of the SiS abundance with respect to H$_2$, already observed between 5 and 20\rstar{} \citep*[e.g.,][]{boyle_1994,agundez_2012,fonfria_2015}.
Both the strength of the pumping emission and the MIR opacity of the SiS in the maser emitting region decrease significantly diminishing at the same time the density of the SiS molecules with inverted levels.
On the other hand, even if the shell thickness is not well determined, this structure is expected to be hollow because our observations are not compatible with a filled spherical maser emitting region (see below for a physical explanation of this reasoning).

\subsubsection{Compact maser emitting regions}

\begin{figure}
\centering
\includegraphics[width=0.475\textwidth]{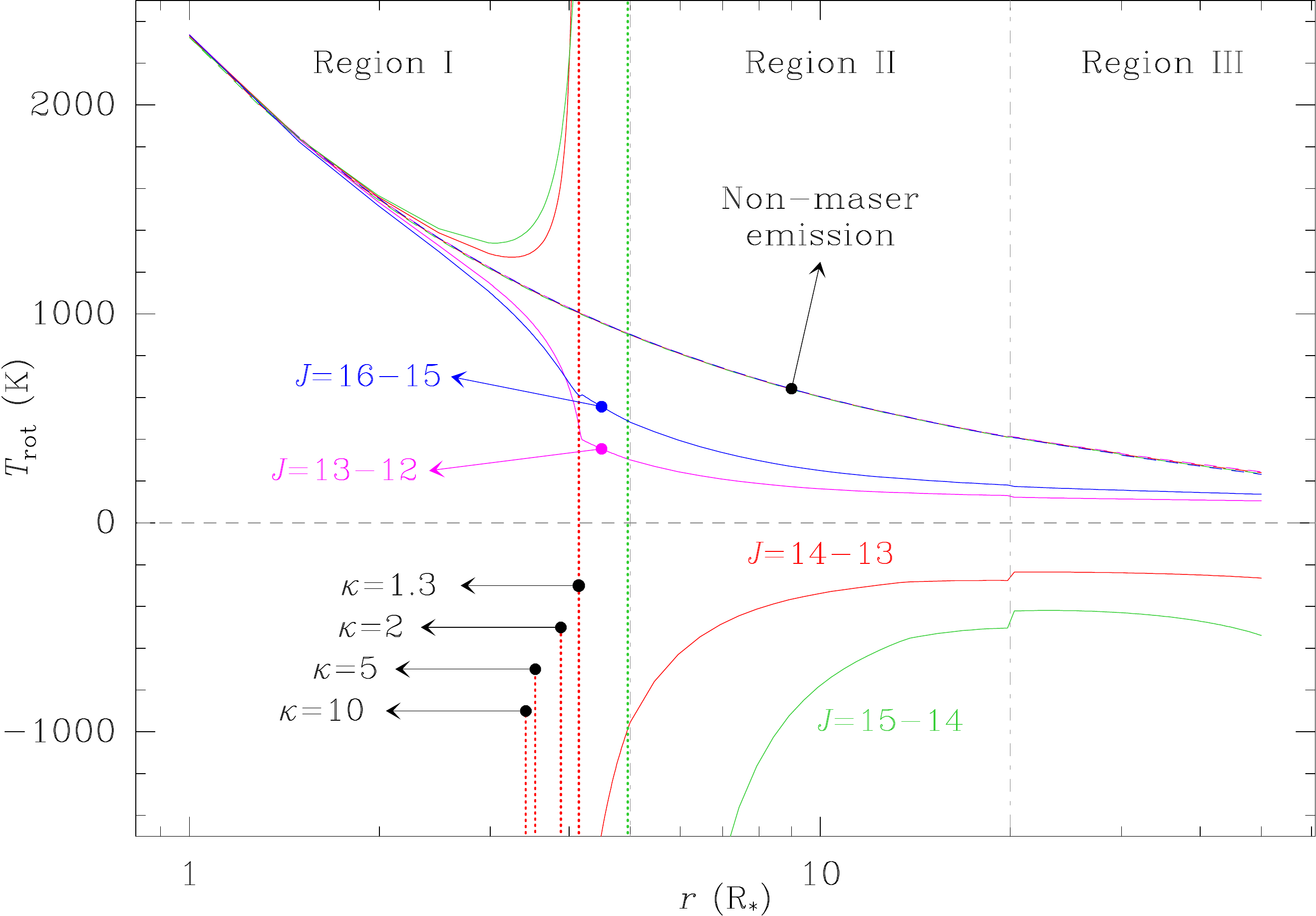}
\caption{Rotational temperature of the SiS lines $v=0$ $J=13-12$, $14-13$, $15-14$, and $16-15$ (magenta, red, green, and blue) in the envelope of \irc{} as a function of the distance from the star.
  The solid curves are the temperatures assuming an enhancement of the continuum intensity for the SiS ro-vibrational transition $1-0$R$_e(9)$, $\kappa$, of a factor of 1.3.
  The dashed curves (all of them superimposed) are the temperatures if this enhancement does not exist (no pumping radiation, $\kappa=0$).
  The dotted vertical lines indicate the distance from the star where the population inversions for the SiS lines $J=14-13$ and $15-14$ are reached for $\kappa=1.3$, 2, 5, and 10 (red and green, respectively; only shown for $J=15-14$ for $\kappa=1.3$).
}
\label{fig:f11}
\end{figure}

Although the presence of the extended emission can be assumed as a direct consequence of the pumping mechanism working in a radially expanding gas envelope, it does not explain so clearly the presence of the derived compact maser structures (the clumps and the arc, which can also be understood as a group of very close clumps).
These clumps are probably related to small regions of the envelope where the pumping mechanism is the same than in the structure that produces the extended maser emission.
However, the maser amplification in them is enhanced compared to their surroundings.
This can be an effect of a local variation of the gas density/SiS abundance with respect to H$_2$ or of the excitation conditions/overlapping degree between the ro-vibrational lines involved in the pumping mechanism.
It is still unclear what could force these physical parameters to vary but we venture that they might be consequence of (1) the increase of the gas density due to spatial irregularities in the mass-loss rate or the presence of arcs, (2) the increase of the molecular abundances caused by a variation in the gas-phase chemistry related to, e.g., an increment in the dissociating radiation field coming from outside the envelope that could reach its inner shells, or (3) inhomogeneities in the dust grain distribution that reduce the dust opacity favoring the increase of the pumping radiation in the maser emitting region.
These scenarios could be compatible with the complex structure composed of thin arcs found in the envelope of \irc{} \citep{mauron_1999,mauron_2000,leao_2006,cernicharo_2015,decin_2015,quintana-lacaci_2016,guelin_2018}.
These structures are predicted to appear naturally in envelopes surrounding single stars by \citet{woitke_2006} and in binary systems by \citet{mastrodemos_1999}, and/or with the peculiar formation of molecules such as H$_2$O, C$_2$H$_4$, C$_4$H$_2$, and carbon chains close to the star \citep{agundez_2010,agundez_2017,fonfria_2017,fonfria_2018}.

Our model suggests that the only maser emission coming from clumps with impact parameters below $5\rstar$ in every velocity channel map is actually produced in Region II.
In fact, there is no need to include clumps in Region I ($r\lesssim 5\rstar$) to reproduce the observations (Figure~\ref{fig:f10}).
The arguments discussed in the previous paragraph can be used to explain the existence of the clumps in Region II but they are also compatible with the existence of clumps closer to the star, where the line width is high, favoring the radiative amplification due to the overlaps expected to exist between lines formed in different emitting volumes.
This apparent incompatibility disappears if we realize that the pumping mechanism invoked to explain the maser emission of the SiS($14-13$) line is a radiative process that can be collisionally canceled in regions with a high kinetic temperature and density, as in Region I.
In order to prove it, we have solved the statistical equilibrium equations for SiS at different distances from the star assuming the physical conditions in \irc{} derived by \citet{fonfria_2015}.
The intensity of the pumping radiation associated to the overlap between the SiS line $1-0$R$(13)$ and the C$_2$H$_2$ line $\nu_5$R$_e(9)$ was modeled as an enhancement of the continuum intensity that connects the SiS levels $v=0$ $J=13$ with $v=1$ $J=14$, keeping untouched the intensity at the frequency of any other line.
The results of this analysis are graphically included in Figure~\ref{fig:f11}, where we show the populations of some rotational levels of the vibrational ground state calculated with and without the continuum intensity enhancement.
We conclude from these results that the rotational excitation of SiS can be considered to be dominated by collisions inwards from $\simeq 3\rstar$, disabling the radiative population inversion mechanism and preventing the maser radiation emission.
This effect explains the hollowness of the extended maser emission inferred from our analysis of the CARMA observations.
From 3 to 5\rstar{} the populations depart from thermal equilibrium as the collisions cannot counteract the effect of the pumping radiation, getting inverted eventually for the levels $J=13$, 14, and 15.
The position in the envelope where the population inversion is reached depends on the intensity of the pumping radiation: the stronger this intensity, the closer to the star the population inversion occurs.

Figure~\ref{fig:f11} shows that it is in principle possible to find maser emission beyond $\simeq 20\rstar$ in disagreement with the results derived from the modeling of our observations.
Moreover, the population inversion in Region III might even be boosted by the decrease of the gas density due to the gas acceleration occurring around $20\rstar$, which weakens the effect of collisions on the level populations.
However, the incompatibility between theory and observations is actually apparent and it can be explained by the gas kinematic properties around $\simeq 20\rstar$.
The turbulent velocity in this region of the envelope is as low as $\simeq 1$~\kms{} and the expected acceleration of $\simeq 3-4$~\kms{} \citep{huggins_1986,keady_1988,skinner_1999,fonfria_2008,fonfria_2015,agundez_2012} could disable the pumping mechanism that we propose to invert the populations, which is very sensitive to frequency shifts.
In this situation, the overlap in the MIR between the C$_2$H$_2$ line coming from Region II, where the bulk of the pumping radiation comes from, and the SiS line in Region III is significantly reduced strongly hampering the population inversion of the SiS rotational levels.
Additionally, we also need to have in mind that the fact that the populations of two levels are inverted does not necessarily mean that we could observe maser emission.
The population inversion have to be strong enough to produce noticeable effects.
Otherwise, its effects could be negligible or be confused with suprathermal emission.

\subsection{Time variability}
\label{sec:discussion.variability}

As we noted in Section~\ref{sec:time.variability}, not all the SiS emission clearly follows the stellar pulsation.
There are shorter scale variations in several lines (e.g., $J=5-4$ and $6-5$) that in principle cannot be satisfactorily explained by this mechanism.
The radiation field coming from the star and the dusty component of the envelope is expected to be driven by the stellar pulsation that varies periodically with time.
Consequently, the effect of this emission on the molecules in the circumstellar envelope has to be also periodical.
However, short time scale variations of the radiation field due to, for instance, changes of the stellar pulsation, which is not very well known so far, could leave noticeable tracks on the molecular emission.
Moreover, we note that the envelope of \irc{} is a very complicated environment that contains a large number of molecules affecting to each other and unexpected effects could arise if we consider a realistic time dependent envelope model.

The comparison of the SiS($14-13$) and SiS($15-14$) lines acquired with CARMA (Jan 2011), ALMA (Apr 2012), and the IRAM 30~m telescope (Jun 2004 and monitoring program, 2015 to present) reveals that their maser structure comprises two substructures that have different time dependences.
Here we assume that the maser emission of both SiS lines is produced by the same structure because the pumping mechanism involves both lines.
On one hand, the absence/presence of the most blue- and red-shifted maser components (A$^*$ and D$^*$) when lines $J=14-13$ (CARMA) and $15-14$ (ALMA) are compared ($\simeq 1.3$~yr lapse between both observations) indicates significant changes with time (see Figure~\ref{fig:f1}).
This is supported by the fact that these maser components are also present in the IRAM 30~m telescope observations of both lines taken by us in 2004 and in the period from 2015 to present, as well as in the single-dish monitoring of the SiS($14-13$) line recently carried out by \citet{he_2017}.
The time monitorings show the strongest maser peaks appearing and disappearing while the rest of the line profiles stay reasonably constant (Figure~\ref{fig:f6}).
Hence, the formation of components A$^*$ and D$^*$ depends strongly on the pulsation phase.
On the other hand, components A, B, C, and D vary much less with respect to the bulk of the line indicating that they are mildly affected by changes in the stellar emission or the envelope evolution.

One of the most stunning results of our analysis is the existence of the phase difference of $\simeq 0.17$ ($\simeq 110$~days) between the NIR light-curve and that of component A$^*$ with respect to the $J=14-13$ and $15-14$ SiS lines (Figure~\ref{fig:f7}).
The obvious similarity between this dependence and the NIR light-curve denotes that this component is driven by the radiation coming from the central star.
The time variation of other molecular species confirm the existence of this phase difference \citep{pardo_2018}.
    
The observations carried out by \citet{shenavrin_2011} and used by \citet{menten_2012} were taken in bands $J$, $H$, $K$, $L$, and $M$ in the period from 1984 to 2008, covering $\simeq 5.5$ pulsation cycles.
The fit to the observed data is reasonably good and although small departures can be noticed in a couple of cycles, they could only imply phase differences of a few percent with respect to the fitted light-curve.
However, \irc{} can only be observed during part of the year preventing us to trace properly the light-curve  \citep{witteborn_1980,lebertre_1988,lebertre_1992,dyck_1991,monnier_1998,shenavrin_2011}.
Consequently, we do not know neither how the actual light-curve is nor whether this shape is constant over time.
Thus, the assumption of a sine-like dependence on time imposes strong constraints on the reference Julian Date (usually at maximum light), the pulsation period, and the amplitude of the brightness variation, that perhaps do not represent accurately the real pulsation process.
In fact, significant variations of the pulsation period over time, usually accepted as produced by thermal pulses, have been reported for several Mira stars \citep*[][and references therein]{templeton_2005}.
Among them, T UMi is one of the best examples of Mira star showing such a long-term variation with a fast decrease of $\simeq 10$~days every $3-4$ pulsation cycles.
Other stars could even display sudden variations on the pulsation period superimposed to their long-term behavior.
Hence, it is possible that the light-curve has changed from the end of the data acquisition to date.

Nevertheless, we have to consider the possibility that the time dependence molecular excitation can produce unexpected delays in the observed molecular emission.
The radiative and collisional transitions happen in time scales shorter than 20~d for most of them \citep*[$A_{v'\to v'',J'\to J''}\gtrsim 6\times 10^{-7}$~s$^{-1}$ for all the lines but the rotational $J=1-0$;][]{cernicharo_2012} and they are usually much faster.
Only the time scales related to the radiative transitions $J=1-0$ in every vibrational state are longer than the detected phase difference \citep*[$A_{v\to v,J=1\to 0}\simeq 7\times 10^{-8}$~s$^{-1}$;][]{cernicharo_2012}.
Hence, it is unlikely that this offset is a consequence of the time evolution of the populations of the SiS ro-vibrational levels.
However, a time evolution excitation effect cannot be completely ruled out if we consider the whole envelope in the problem.
In particular, IR light could take as long as $\simeq 10-20$~days to reach the external layers where \citet{pardo_2018} have observed the spectra of several molecules showing a phase lag similar to what we have found for SiS.
An accurate description of the molecular emission dependence with time ought to rely on a complex non-local, time-dependent excitation model, something out of the scope of the current work but that we will address in the future.

\section{Summary and conclusions}
\label{sec:summary}

In this paper, we have presented new high angular, high spectral resolution observations of lines SiS($14-13$) and SiS($15-14$) toward the AGB star \irc, acquired with the CARMA and ALMA millimeter arrays.
Our interferometer data show that most of the emission has a maser nature.
We have used a 3D radiation transfer code to model the observed emission distribution of the SiS($14-13$) line in order to propose an envelope model that could explain the observations.
These interferometer data have been complemented with a detailed monitoring of these and other lines of SiS (and $^{29}$SiS) carried out with the IRAM 30~m telescope that covered a whole pulsation period.
After studying all these data sets, we conclude that:

\begin{itemize}

\item A large fraction of the maser emission of the SiS($14-13$) line comes from two compact regions located at $0\farcs07\pm 0\farcs04$ to the NW with respect to the star in front of and behind it as well as other weaker compact sources.
A bipolar structure with a P.A.~$\simeq 80\degr$ peaking at $0\farcs17\pm 0\farcs04$ and $0\farcs13\pm 0\farcs04$ at both sides of the star has been discovered in the brightness distribution coming from gas expanding roughly at the systemic velocity.
  
\item The emission comprises several compact maser structures embedded in extended emission.
The thermal emission cannot explain this extended emission leading us to think that the extended emission is also composed of a significant maser contribution.
  
\item The maser emitting structure can be described by a group of clumps spanning several velocity channels along the line-of-sight and a shell-like structure that accounts for the extended maser emission. 
  One arc roughly contained in the plane of the sky and in front of the star is also needed to reproduce correctly the observations.
  
\item The clumps and the arc are distributed between 5 and 20\rstar.
The clumps responsible for the strongest maser spots are located in the NW quadrant ranging from 5 to 11\rstar.
The extended maser emission can be reproduced with the thermal emission and a homogeneous emitting shell-like region with a radius of $\simeq 13\rstar$ and an effective thickness of $\simeq 2\rstar$ with a variable negative excitation temperature.

\item The maser emitting structures (the clumps, the arc, and the extended maser emission) account for 75\% of the total emission while only 25\% is thermal emission.
About 40\% of the whole maser emission comes from the clumps and the arc.
The extended maser emission is the responsible of most of the maser emission ($\simeq 60$\%).

\item This model is compatible with the pumping mechanism based on overlaps in the MIR between the lines SiS $1-0$R$(13$) and C$_2$H$_2$ $\nu_5$R$_e(9)$ that can occur between gas volumes expanding along different directions or with different velocities.
One of the consequences of this mechanism is the existence of extended maser emission due to the roughly spherical symmetry of the expansion velocity field, something that can explain the observations.
  
\item The monitoring of the SiS $J=14-13$ and $15-14$ lines indicates that some of their narrow maser components show evident variations in time related to the NIR light-curve of star.
Other spectral components display a milder time variability.
The large variations experienced by these lines compared to that of other thermally excited SiS lines suggests that the extended maser emission also varies with time.
  
\item The dependence of the emission related to the strongest one is offset by $\simeq 0.2$ pulsation periods with respect to the NIR light-curve.
\end{itemize}

\section*{Acknowledgements}

We thank all members of CARMA staff that made observations possible.
We also thank the anonymous referee for his/her comments.
During the first part of this work, JPF was supported by the UNAM through a postdoctoral fellowship.
JC, MA, LVP, GQ-L, MS-G, JRP and JPF thank the Spanish MINECO/MICINN for funding support through grants AYA2009-07304, AYA2012-32032, AYA2016-75066-C-1-P, the ASTROMOL Consolider project CSD2009-00038 and the European Research Council (ERC Grant 610256: NANOCOSMOS).
MA also thanks funding support from the Ram\'on y Cajal program of Spanish MINECO (RyC-2014-16277).


\appendix

\section{Calibration issues during the time monitoring}
\label{sec:cal.issues}

Time monitoring of the lines SiS $v=0$ $J=5-4$, $6-5$, $14-13$, and $15-14$, and $^{29}$SiS $v=0$ $J=13-12$ and $15-14$ coming from \irc{} was carried out to study in reasonable detail one pulsation period of the central star (see Figure~\ref{fig:f6}).
Reliable absolute calibration, focusing, and pointing through the monitoring is much easier to achieve for the lowest frequencies ($80-115$~GHz band) whereas errors related to these processes dramatically increase with frequency.
After checking the effect of these calibration issues on the high frequency data, we estimate flux uncertainties of $15-20\%$ for the SiS and $^{29}$SiS lines.
Nevertheless, since the source is much smaller than the antenna main beam at all frequencies, the shape of the line should not suffer significantly from these uncertainties and, therefore, the time evolution of the line profiles should be physically meaningful.
However, the line fluxes can be affected by these unavoidable errors (mostly at high frequencies), arising incompatibilities between consecutive observations that would mask the possible time dependence of the thermal and extended maser emissions.

We have limited the effect of the calibration uncertainties in the 1~mm band with the aid of auxiliary thermal lines of molecular species such as HNC or SiC$_2$, which emission is known to be quite constant \citep{cernicharo_2014}.
These lines were observed at the same time and with the same setup than the SiS and the $^{29}$SiS lines.
We have used these auxiliary lines to recalibrate the SiS and $^{29}$SiS lines at high frequency.
Under these circumstances, any sudden variation in the SiS and $^{29}$SiS is expected to be real and could be directly associated to the fast changes expected to occur in systems very sensitive to the pumping mechanism such as the maser clumps.

\section{Considerations on the code and the modeling process}
\label{sec:considerations}

The approach adopted to estimate the level populations with our code allows us to provide negative excitation temperatures.
Contrarily to the procedure followed to calculate the thermal emission, which assumes that the rotational and vibrational temperatures throughout the envelope follow
a power-law ($\propto r^{-\gamma}$), the excitation temperature profile within maser emitting regions is built by linearly interpolating the input values of this temperature at several selected points.
The transition of the excitation temperature between adjacent thermal and maser emitting shells (positive to negative and vice versa) is described taking advantage of 
the continuity of the level populations, related to the excitation temperature through a Boltzmann factor.
The maser emission is assumed to grow exponentially with the optical depth, which is particularly suitable for the treatment of weak masers.

The modeling of the maser emission of any of these lines is an ambitious task because of the inaccuracies inherent to the numerical treatment of the growing exponential dependence on the optical depth of the maser intensity.
Assuming a fixed grid to calculate the emission in the maser emitting regions is inaccurate due to the unknown position of the emission peak in the defined grid.
This issue results in significant variations of the synthetic brightness distribution between runs with slightly different conditions, suggesting a strong and frequently unexpected dependence of the maser emission on the model and the number of points of the grid.
This problem could be solved by a massive refining of the grid around the maser emitting regions, which increases the CPU time substantially.
Thus, we decided to define a pseudo-random grid only in the maser emitting regions.

Selecting a reasonable high number of points in this pseudo-random grid and repeating several times the calculations, we were able to estimate the emitting flux within a statistical interval.
Achieving variations similar to the observed noise RMS could require a prohibitive CPU time to find the set of parameters that best reproduce the observations so that a compromise between the synthetic emission accuracy and the time devoted to perform the modeling needs to be reached.
In our case, the deviation from the synthetic to the observed emission has been adopted to be up to $\simeq 10$\% the peak emission in each velocity-channel map ($\simeq 1.5$~\jybeam).
Despite these deviations are much higher than the observed noise RMS ($\sigma_\subscript{noise}\simeq 45$~m\jybeam), it is enough to derive reliable results regarding the main components of the brightness distribution that show peak values higher than 150~m\jybeam{} ($\simeq 3\sigma_\subscript{noise}$).

We use the same method followed by \citet{fonfria_2014} to compare the synthetic emission with the observed one.
After the model has been calculated, it is possible to replace the observed visibilities contained in the original Miriad packages by those derived from the model with the Miriad task \textsc{uvmodel}.
This action let us to apply the same constraints on our model than the array on the actual brightness distribution.
Thus, we can directly compare the synthetic emission with the observations since the same procedure is used to map the observed data and the model beginning from the visibilities inversion.
Part of the synthetic emission is also filtered by the lack of short baselines.

\end{document}